\documentclass{aa}
\usepackage{graphicx}
\usepackage{bm}
\usepackage{times}
\usepackage{amsmath}
\usepackage{amssymb}
\usepackage{natbib}
\usepackage{color}
\usepackage{multirow}
\usepackage{rotating}
\bibpunct{(}{)}{;}{a}{}{,}
\graphicspath{{./fig/}{./png/}}
\newcommand{\bbb}{{\bm b}}

\newcommand{\uuu}{{\bm u}}

\newcommand{\UUU}{{\bm U}}

\newcommand{\BBB}{{\bm B}}

\newcommand{\Eq}[1]{Eq.~(\ref{#1})}
\newcommand{\EQ}{\begin{equation}}
\newcommand{\EN}{\end{equation}}
\newcommand{\EQA}{\begin{eqnarray}}
\newcommand{\ENA}{\end{eqnarray}}
\newcommand{\brac}[1]{\langle #1 \rangle}
\newcommand{\pd}{\partial}

\newcommand{\DIV}{\vec{\nabla} \cdot }

\newcommand{\mean}[1]{\overline{#1}}
\newcommand{\meanv}[1]{\overline{\bm #1}}
\newcommand{\cP}{c_{\rm P}}
\newcommand{\cV}{c_{\rm V}}

\newcommand{\nut}{\nu_{\rm t}}

\newcommand{\urms}{u_{\rm rms}}

\newcommand{\kf}{k_{\rm f}}

\newcommand{\chiSGS}{\chi_{\rm SGS}}
\newcommand{\chiSGSm}{\chi^{\rm m}_{\rm SGS}}

\newcommand{\Co}{{\rm Co}}

\newcommand{\Pe}{{\rm Pe}}
\newcommand{\Pra}{{\rm Pr}}
\newcommand{\PraSGS}{{\rm Pr}_{\rm SGS}}
\newcommand{\PrM}{{\rm Pr}_{\rm M}}
\newcommand{\Pm}{{\rm Pm}}

\newcommand{\Ra}{{\rm Ra}}

\newcommand{\Rey}{{\rm Re}}

\newcommand{\Rm}{{\rm Rm}}
\newcommand{\ReM}{{\rm Re}_{\rm M}}
\newcommand{\Ro}{{\rm Ro}}

\newcommand{\Ta}{{\rm Ta}}

\newcommand{\qqrr}{\mathcal{Q}_{rr}}
\newcommand{\qqtt}{\mathcal{Q}_{\theta\theta}}
\newcommand{\qqpp}{\mathcal{Q}_{\phi\phi}}

\newcommand{\qqrp}{\mathcal{Q}_{r\phi}}
\newcommand{\qqtp}{\mathcal{Q}_{\theta\phi}}

{}
\def\onethird{{\textstyle{1\over3}}}
\def\onehalf{{\textstyle{1\over2}}}

\newcommand{\Fig}[1]{Fig.~\ref{#1}}
\newcommand{\Sec}[1]{Sect.~\ref{#1}}
\newcommand{\Table}[1]{Table~\ref{#1}}
\begin{document}

\authorrunning{K\"apyl\"a et al.}
\titlerunning{Convection-driven spherical shell dynamos at varying Prandtl numbers}

   \title{Convection-driven spherical shell dynamos\\ at varying Prandtl numbers}

   \author{P. J. K\"apyl\"a
          \inst{1,2,3,4}
          \and
          M. J. K\"apyl\"a
          \inst{3,2}
          \and
          N. Olspert
          \inst{2}
          \and
          J. Warnecke
          \inst{3}
          \and
          A. Brandenburg
          \inst{4,5,6,7}
          }

   \institute{Leibniz-Institut f\"ur Astrophysik Potsdam, 
              An der Sternwarte 16, D-11482 Potsdam, Germany
              \email{pkapyla@aip.de}
         \and ReSoLVE Centre of Excellence, Department of Computer Science,
              Aalto University, PO Box 15400, FI-00076 Aalto, Finland
         \and Max-Planck-Institut f\"ur Sonnensystemforschung,
              Justus-von-Liebig-Weg 3, D-37077 G\"ottingen, Germany
         \and NORDITA, KTH Royal Institute of Technology and Stockholm University,
              Roslagstullsbacken 23, SE-10691 Stockholm, Sweden
         \and Department of Astronomy, AlbaNova University Center,
              Stockholm University, SE-10691 Stockholm, Sweden
         \and JILA and Department of Astrophysical and Planetary Sciences,
              Box 440, University of Colorado, Boulder, CO 80303, USA
         \and Laboratory for Atmospheric and Space Physics,
              3665 Discovery Drive, Boulder, CO 80303, USA
}

\date{\today,~ $ $Revision: 1.287 $ $}

\abstract{Stellar convection zones are characterized by
   vigorous high-Reynolds number turbulence at low Prandtl numbers.
}{%
  We study the dynamo and differential rotation regimes at varying levels
   of viscous, thermal, and magnetic diffusion.
}%
   {
   We perform three-dimensional simulations of stratified fully
   compressible
   magnetohydrodynamic convection in rotating
   spherical wedges at various thermal and magnetic Prandtl numbers
   (from 0.25 to 2 and 5, respectively).
   Differential rotation and large-scale magnetic
   fields are produced self-consistently.
   }%
   {
   We find that for high thermal diffusivity, the rotation profiles show
   a monotonically increasing angular velocity from the bottom
   of the convection zone to the
   top and from the poles toward the equator.
   For sufficiently rapid rotation, a region of
   negative radial shear develops at mid-latitudes as the thermal diffusivity is
   decreased, corresponding to an increase of the Prandtl number. This
   coincides with and results in a change of the dynamo mode
   from poleward propagating
   activity belts to equatorward propagating ones. Furthermore, the
   clearly cyclic solutions disappear at the highest magnetic Reynolds
   numbers and give way to irregular sign changes or quasi-stationary states.
   The total (mean \& fluctuating) magnetic energy increases as a function of
   the magnetic Reynolds number in the range studied here ($5-151$), but the
   energies of the mean magnetic fields level off at high magnetic Reynolds
   numbers. The differential rotation is strongly affected by the
   magnetic fields and almost vanishes at the highest magnetic
   Reynolds numbers.
   In some of our most turbulent cases, however, we find that two
   regimes are possible where either differential rotation is strong
   and mean magnetic fields relatively weak or vice versa.
   }%
   {
   Our simulations indicate a strong non-linear feedback of magnetic
   fields on differential rotation, leading to qualitative changes in
   the behaviors of large-scale dynamos at high magnetic Reynolds
   numbers. Furthermore, we do not find indications of the simulations
   approaching an asymptotic regime where the results would be
   independent of diffusion coefficients in the parameter range
   studied here.
   }%


  \maketitle


\section{Introduction}

Simulations of convection-driven dynamos have recently reached a level
of sophistication where they capture effects observed in the Sun such
as equatorward migration of activity belts
\citep{SPD11,KMB12,ABMT15,DWBG16}, and
irregular cycle variations such as grand minima and long term
modulations \citep{PC14,KKOBWKP16}. Most of these simulations are
individual numerical experiments and it is not clear how they are
situated in parameter space in relation to each other.
Important parameters in this connection concern the relative strengths of
different diffusion coefficients, viscosity ($\nu$), magnetic
($\eta$), and thermal ($\chi$) diffusivities present in the system.
Their ratios are characterized by the thermal and magnetic \emph{Prandtl numbers},
$\Pra=\nu/\chi$ and $\PrM=\nu/\eta$, respectively.
In the solar convection zone, these Prandtl numbers are $\Pra\ll1$ and $\PrM\ll1$, while the
fluid and magnetic Reynolds numbers, $\Rey=ul/\nu$ and $\ReM=ul/\eta$,
with $u$ and $l$ being the characteristic velocity and length scale, are
of the orders of $10^{12}$ and $10^9$, respectively. Such parameter
regimes are not accessible to current numerical simulations which are
restricted to $\Pra\approx1$, $\PrM\approx1$, and Reynolds numbers of
the order to $10^2\ldots10^3$.
In all simulations by a number of different groups,
the dominant contribution to thermal diffusion
comes through a subgrid-scale (SGS) coefficient
$\chiSGS$ whose magnitude is much higher than the radiative one. 
Similar arguments apply also to $\nu$ and $\eta$, but since the
functional form of those diffusion operators is unchanged,
we omit in those the subscript SGS.
Thus, the relevant thermal Prandtl number in
simulations is $\PraSGS=\nu/\chiSGS$.
We emphasize that this applies to simulations of all groups,
although the nomenclature may be different (see \Table{tab:methods} in
Appendix~\ref{sec:comp}).
This is also true for groups using realistic luminosities, and thus the
correct order of magnitude for the radiative diffusivity
\citep[e.g.][]{BMT04,HRY16}.

When the convection simulations of \cite{KMB12} in wedge geometry showed
for the first time equatorward migration, it was not yet clear
that this was related to their choice of $\PraSGS=2.5$ compared
with $\PraSGS \lesssim 1$ used in most earlier simulations that
showed either quasi-stationary configurations \citep{BBBMT10} or
either weak or poleward migration
\citep[e.g.][]{GCS10,KKBMT10,BMBBT11,GDW12}. Recently, \cite{WKKB14} showed that
the change in the dynamo behavior between $\PraSGS>1$ and $\PraSGS<1$
regimes is due to a change in the differential rotation profile
which, in the $\PraSGS\gtrsim1$ regime, leads to a region of negative radial
shear that facilitates the equatorward migration.
Also the magnetic Prandtl number, which is proportional to the
magnetic Reynolds number, can strongly affect the results.
Increasing $\ReM$ by increasing $\PrM$
can allow magnetohydrodynamic (MHD) instabilities such
as magnetic buoyancy \citep{Pa55a} and magnetorotational instabilities
\cite[e.g.][]{Pm07,Ma11} to be excited. 

Increasing the magnetic
Reynolds number can influence the large-scale dynamo via several other
avenues. First, the most easily excited dynamo mode can
change. Second, a small-scale dynamo is likely excited after $\ReM$
exceeds a threshold value \citep[e.g.][]{Cat99}, and this may also
affect the large-scale dynamo by modifying the velocity field. Third,
Boussinesq
simulations indicate that differential rotation is strongly quenched
as the magnetic Reynolds number increases \citep{SPD12}. This was
shown to be associated with a transition from oscillatory multipolar
large-scale field configurations to quasi-stationary dipole-dominated
dynamos as a function of $\ReM$. One
of the main goals of the present paper is therefore to systematically study the
effects of varying Prandtl numbers on the differential rotation and
dynamo modes excited in the simulations.
We note that similar parameter studies have been performed with
Boussinesq simulations \citep[e.g.][]{SB05,BS06}. Here we explore the
stratified, fully compressible simulations and reach parameter regimes
that are significantly more supercritical in terms of both the convection
and the dynamo.

Another important aspect is related to the saturation level of the
large-scale field in simulations at high $\ReM$. Dynamo theory experienced a crisis
in the early 1990s when it was discovered that the energy of the large-scale
magnetic field saturates at a level that is inversely proportional to the
magnetic Reynolds number
\citep{GD95,BD01}. If this were to carry over to the Sun, where
the magnetic Reynolds number is of the order of $10^9$ or greater, only
very weak large-scale fields would survive. This phenomenon was related to a
\emph{catastrophic quenching} of the $\alpha$ effect \citep{CV91,VC92,CH96}.

Later, this was understood in terms of magnetic helicity: if the system
is closed or fully periodic, i.e.\ when magnetic field lines do not
cross the boundary of the system, no flux of magnetic helicity in or
out can occur, and only the \emph{molecular} diffusion can change it
\citep{B01}. In astrophysical systems this would mean that magnetic
helicity would be nearly conserved. However, astrophysical systems are
not closed and magnetic helicity can escape, e.g.\ via coronal mass
ejections in the Sun \citep[e.g.][]{BB03,WBM11,WBM12} or via winds from
galaxies \citep{SSSB06,SSS07,DSGB13}. 
In mean-field theory
these physical effects are parameterized by fluxes, which lead to
alleviation of catastrophic quenching in suitable parameter regimes
\citep[e.g.][]{BCC09}.

Direct numerical simulations of large-scale dynamos have demonstrated
that open boundaries lead to alleviation of catastrophic quenching in
accordance with the interpretation in terms of magnetic helicity
conservation \citep[e.g.][]{BS04,KKB10b}. Although the large-scale
magnetic field amplitude does not decrease proportional to $\ReM$ in the
cases when open boundaries are used, there is still a decreasing trend
even at the highest currently studied $\ReM$
in local simulations of convection-driven dynamos
\citep[e.g.][]{KKB10b}. This, however, is compatible with mean-field
models which suggest that the magnetic
helicity fluxes become effective only at significantly higher $\ReM$
\citep{BCC09,DSGB13}.

In convective dynamos in spherical coordinates the
computational challenge is even greater and systematic parameter scans
have not been performed.
Some preliminary attempts have been made, but the results remain inconclusive.
An illuminating example is the study of
\cite{NBBMT13} where the large-scale axisymmetric field decreases by a
factor of two when the magnetic Reynolds number is increased by a
factor of four, which is still rather steep. In a recent paper,
\cite{HRY16} show that in even higher-$\ReM$ simulations the mean
magnetic energy recovers and claim that this is a consequence of an
efficient small-scale dynamo that suppresses small-scale flows.
Another goal of the present paper
is therefore to study the saturation level of the large-scale
field in convection-driven dynamos in spherical coordinates with and
without a simultaneous small-scale dynamo (hereafter SSD).

In the present study, we employ a spherical wedge geometry by imposing
either a perfect conductor or a normal field boundary condition at
high latitudes.
Earlier mean-field simulations of $\alpha\Omega$ dynamos have suggested
that solutions with a perfect conductor boundary condition are similar
to those in full spherical shells \citep{JBTM90}.
However, more recent work by \cite{CBKK16} has demonstrated that this
conclusion is not generally valid and depends on the nature of the
solutions.
Their work also suggests that the use of a normal field boundary condition
at high latitudes might be a better way of obtaining solutions that are
applicable to full spherical shells.
Owing to this uncertainty, we investigate here cases with both types of
boundary conditions.

\section{The model} \label{sect:model}

The model is similar to that used in \cite{KMB12} and is described in
detail in \cite{KMCWB13}.
We study the
dynamics of magnetized gas in spherical coordinates where only parts
of the latitude and longitude ranges are retained.
More specifically, we model a wedge that covers $r_0\le r\le r_1$ in
radius, $\theta_0\le\theta\le\pi-\theta_0$ in colatitude, and
$0\le\phi\le\phi_0$ in longitude. Here we use $r_0=0.7R_\odot$,
$r_1=R_\odot$, and where $R_\odot=6.96\cdot10^8$m is the solar radius,
$\theta_0=15\degr$, and $\phi_0=90\degr$.

We solve the following set of compressible hydromagnetics equations
\begin{eqnarray}
\frac{\pd {\bm A}}{\pd t} &=& {\bm U} \times {\bm B} - \eta \mu_0 {\bm J},\\
\frac{D \ln \rho}{Dt} &=& - \bm\nabla\bm\cdot{\bm U}, \\
\frac{D {\bm U}}{Dt} &=& {\bm g}  - 2\bm\Omega_0\times \bm U + \frac{1}{\rho} [\bm\nabla\!\bm\cdot\!(2\nu\rho \bm{\mathsf{S}}) - \bm\nabla p + {\bm J} \times {\bm B}], \\
T \frac{Ds}{Dt} &=& \frac{1}{\rho} \left[\eta \mu_0 {\bm J}^2-\bm\nabla\bm\cdot({\bm F}^{\rm rad}+{\bm F}^{\rm SGS})\right]+2\nu \bm{\mathsf{S}}^2,
\label{equ:1}
\end{eqnarray}
where ${\bm A}$ is the magnetic vector potential, ${\bm U}$ is the
velocity, ${\bm B} = \bm\nabla\times{\bm A}$ is the magnetic field,
$\eta$ is the magnetic diffusivity, $\mu_0$ is the permeability of
vacuum, ${\bm J}=\bm\nabla\times{\bm B}/\mu_0$ is the current density,
$D/Dt = \pd/\pd t + {\bm U}\bm\cdot\bm\nabla$ is the advective time
derivative, $\rho$ is the density, ${\bm g}$ is the acceleration due
to gravity and $\bm\Omega_0=(\cos\theta,-\sin\theta,0)\Omega_0$ is
the angular velocity vector where $\Omega_0$ is the rotation rate of
the frame of reference,
$\nu$ is the kinematic viscosity, $p$ is the pressure, and
$s$ is the specific entropy with $Ds=\cV D\ln p-\cP D\ln\rho$.
The gas is assumed to obey the ideal gas
law, $p=(\gamma-1) \rho e$, where $e=\cV T$ is the specific internal
energy and $\gamma=\cP/\cV$ is the ratio of specific heats at constant
pressure and volume, respectively. The rate of strain tensor is given by
\begin{eqnarray}
\mathsf{S}_{ij} = \onehalf (U_{i;j} + U_{j;i}) - \onethird \delta_{ij} \bm\nabla\bm\cdot {\bm U},
\end{eqnarray}
where the semicolons refer to covariant derivatives; see
\cite{MTBM09}. The radiative and subgrid-scale (SGS) fluxes are given
by
\begin{eqnarray}
{\bm F}^{\rm rad} = -K\bm\nabla T, \hspace{1cm }{\bm F}^{\rm SGS} = -\chiSGS \rho T \bm\nabla s,
\label{equ:Frad}
\end{eqnarray}
where $K=\cP \rho \chi$ is the heat conductivity, and $\chiSGS$ is the
(turbulent) subgrid-scale diffusion coefficient for the entropy.

\subsection{Initial and boundary conditions}
We follow the description given in \cite{KKB14} to transform our
results into physical units. As our equations are fully compressible
we cannot afford to use the solar luminosity which would lead to the
sound speed dominating the time step. Thus we increase the luminosity
substantially and to compensate we increase the rotation rate to
achieve the same rotational influence as in the Sun. Assuming a
scaling of the luminosity with the convective energy flux as $L
\propto \rho u^3$, we find that the convective velocities increase to
one third power of the luminosity \citep[e.g.][]{BCNS05,KKKBOP15}. The
ratio between model and solar
luminosities is $L_0/L_\odot\approx6.4\cdot10^5$ in the
current simulations. 
We correspondingly increase the rotation rate by a factor of
$(L_0/L_\odot)^{\frac{1}{3}}\approx 86$ for the solar case. We
reiterate that the main effect of increasing the luminosity is to
increase the Mach number and bring the acoustic and dynamical
timescales closer to each other, which facilitates the computations
with a fully compressible method \citep[cf.\ Fig.\ 1
  of][]{KMCWB13}. Another possibility would be to apply the so-called
reduced sound speed method where the sound speed is artificially
changed so that the Mach number does not become too small at the base
of the convection zone \citep{HRYIF12,KBKKR16}.

The higher luminosity also helps to reach a thermal equilibrium within
reasonable simulation running time.
Furthermore, we assume that the
density and the temperature at the base of the convection zone are the
same as in the Sun, i.e.\ $\rho_{\rm bot}=200$~kg~m$^{-3}$ and
$T=2.23\cdot10^6$~K.

As the initial condition for the hydrodynamics we use an isentropic
atmosphere. The temperature gradient is given by
\begin{equation}
\frac{\pd T}{\pd r} = -\frac{GM_\odot/r^2}{\cV(\gamma-1)(n_{\rm ad}-1)},
\end{equation}
where $G=6.67\cdot10^{-11}$~N~m$^2$~kg$^{-2}$ is Newton's
gravitational constant, $M_\odot=1.989\cdot10^{30}$~kg is the solar
mass, and $n_{\rm ad}=1.5$ is the polytropic index for an adiabatic
stratification.

The initial state is not in thermal equilibrium but closer to the
final convecting state to reduce the time needed to reach a
statistically stationary state. The heat conductivity has a profile
given by $K(r)=K_0[n(r)+1]$, where $n(r)=\delta n (r/r_0)^{15} +
n_{\rm ad} -\delta n$, with
$K_0=(\mathcal{L}/4\pi)\cV(\gamma-1)(n_{\rm ad}+1)\rho_{\rm bot}
\sqrt{GM_\odot R_\odot}$,
and where $\mathcal{L}$ is a dimensionless luminosity defined below.
We keep $\delta n=1.9$ fixed in our simulations resulting a situation
where radiation transports all of the flux into the domain but its
contribution diminishes rapidly toward the surface \citep[see,
  e.g., Fig.~2 of][]{KMB11}.

The subgrid scale diffusivity $\chiSGS$ for the entropy has a
piecewise constant profile, such that the value
$\chiSGS=6.1\cdot10^8$~m~s$^{-2}$ is fixed above $r=0.98R_\odot$ in
all runs, and the value $\chiSGSm=\chiSGS(r=r_m=0.85R_\odot)$ in the
bulk of the convection zone 
is varied via the corresponding Prandtl number (see below). The value
below $r=0.75R_\odot$ is set equal to $0.1\chiSGSm$.
The constant values in the different layers connect smoothly 
over a transition depth of $d=0.01R_\odot$.

The boundary conditions for the flow are assumed to be impenetrable and
stress free, i.e.,
\begin{eqnarray}
&&\!\!\!
U_r=0,\quad \frac{\pd U_\theta}{\pd r}=\frac{U_\theta}{r},\quad \frac{\pd
U_\phi}{\pd r}=\frac{U_\phi}{r} \quad (r=r_0, r_1),\\
&&\!\!\!
\frac{\pd U_r}{\pd \theta}=U_\theta=0,\quad \frac{\pd U_\phi}{\pd
\theta}=U_\phi \cot \theta \quad (\theta=\theta_0,\pi-\theta_0).
\quad
\end{eqnarray}
The lower radial boundary is assumed to be perfectly conducting, and
on the outer radial boundary the field is purely radial. The latitudinal
boundaries are either perfectly conducting (PC) or a normal field (NF)
condition is assumed. In terms of the magnetic vector potential these
are given by
\begin{eqnarray}
&&\!\!\!
\frac{\pd A_r}{\pd r}= A_\theta=A_\phi =0 \,\quad
(r=r_0),\\
&&\!\!\!
A_r=0, \;\; \frac{\pd A_{\theta}}{\pd r}=-\frac{A_{\theta}}{r},\;\; \frac{\pd
A_{\phi}}{\pd r}=-\frac{A_{\phi}}{r} \quad (r=r_1),\quad\\
&&\!\!\!
A_r=\frac{\pd A_\theta}{\pd\theta}=A_\phi=0 \quad
(\theta=\theta_0,\pi-\theta_0) \quad \mbox{(PC)},\\
&&\!\!\!
\frac{\pd A_r}{\pd\theta}\!=\!A_\theta\!=\!0, \frac{\pd A_\phi}{\pd\theta}\!=\!-\frac{\cot\theta A_\phi}{r}\ 
(\theta=\theta_0,\pi-\theta_0) \ \mbox{(NF)}.
\end{eqnarray}
For the density and specific entropy we assume vanishing first
derivatives on the latitudinal boundaries.

At the lower boundary we specify
\begin{equation}
F_r^{\rm rad}+F_r^{\rm SGS}=\frac{L_0}{4\pi r_0^2},
\end{equation}
which leads to constant input luminosity into the system.
On the outer radial boundary we apply a radiative boundary condition
\begin{equation}
\sigma T^4 = F_r^{\rm rad}+F_r^{\rm SGS},
\label{eq:bbb}
\end{equation}
where $\sigma$ is the Stefan--Boltzmann constant. We use a modified
value for $\sigma$ that takes into account that the luminosity
and the temperature at the surface are larger than
in the Sun. The value of $\sigma$ is chosen so that the surface flux,
$\sigma T^4$, carries the total luminosity through the boundary in the
initial non-convecting state.

\begin{table*}[t!]
\centering
\caption[]{Summary of the runs.}
  \label{tab:runs}
      $$
          \begin{array}{p{0.05\linewidth}ccccccccccccccc}
          \hline
          \hline
          \noalign{\smallskip}
          Run & \tilde\Omega & \Ra [10^6] & \Pra & \PraSGS & \PrM & \Ta [10^8] & \nu [10^8$m$^2$s$^{-1}] & \Rey & \Pe & \ReM & \Co & \Delta t & {\rm SSD} & \mbox{BCs} & \mbox{Grid} \\
          \hline
          A1 &  5 &   0.8 & 71.7 & 0.25 & 0.25 &  1.25 &  1.01 &  20 &   5 &   5 & 14.0 & 0.4 & - &\mbox{PC} & 128 \times 256 \times 128 \\ 
          A2 &  5 &   0.8 & 71.7 & 0.25 & 0.50 &  1.25 &  1.01 &  18 &   4 &   9 & 15.1 & 0.4 & - &\mbox{PC} & 128 \times 256 \times 128 \\ 
          A3 &  5 &   0.8 & 71.7 & 0.25 & 1.00 &  1.25 &  1.01 &  17 &   4 &  17 & 15.9 &  56 & - &\mbox{PC} & 128 \times 256 \times 128 \\ 
          A4 &  5 &   0.8 & 71.7 & 0.25 & 2.00 &  1.25 &  1.01 &  22 &   5 &  45 & 12.4 &  37 & - &\mbox{PC} & 128 \times 256 \times 128 \\ 
          A5 &  5 &   0.8 & 71.7 & 0.25 & 3.33 &  1.25 &  1.01 &  23 &   5 &  79 & 11.7 &  26 & - &\mbox{NF} & 256 \times 512 \times 256 \\ 
          \hline
          B1 &  5 &   2.9 & 71.7 & 0.50 & 0.25 &  1.25 &  1.01 &  25 &  12 &   6 & 11.1 & 5.5 & - &\mbox{PC} & 128 \times 256 \times 128 \\ 
          B2 &  5 &   2.9 & 71.7 & 0.50 & 0.50 &  1.25 &  1.01 &  24 &  12 &  12 & 11.3 &  27 & - &\mbox{PC} & 128 \times 256 \times 128 \\ 
          B3 &  5 &   2.9 & 71.7 & 0.50 & 1.00 &  1.25 &  1.01 &  27 &  13 &  27 & 10.3 &  26 & - &\mbox{PC} & 128 \times 256 \times 128 \\ 
          B4 &  5 &   2.9 & 71.7 & 0.50 & 2.00 &  1.25 &  1.01 &  29 &  14 &  58 &  9.7 &  50 & - &\mbox{PC} & 128 \times 256 \times 128 \\ 
          B5 &  5 &   2.9 & 71.7 & 0.50 & 5.00 &  1.25 &  1.01 &  27 &  13 & 137 & 10.3 &  23 & + &\mbox{NF} & 256 \times 512 \times 256 \\ 
          \hline
          C1 &  5 &    10 & 71.7 & 1.00 & 0.25 &  1.25 &  1.01 &  28 &  28 &   7 &  9.9 &  27 & - &\mbox{PC} & 128 \times 256 \times 128 \\ 
          C2 &  5 &    10 & 71.7 & 1.00 & 0.50 &  1.25 &  1.01 &  29 &  29 &  14 &  9.6 &  44 & - &\mbox{PC} & 128 \times 256 \times 128 \\ 
          C3 &  5 &    10 & 71.7 & 1.00 & 1.00 &  1.25 &  1.01 &  29 &  29 &  29 &  9.4 & 108 & - &\mbox{PC} & 128 \times 256 \times 128 \\ 
          C4 &  5 &    10 & 71.7 & 1.00 & 1.43 &  1.25 &  1.01 &  30 &  30 &  44 &  9.1 &  65 & - &\mbox{PC} & 128 \times 256 \times 128 \\ 
          C5 &  5 &    10 & 71.7 & 1.00 & 5.00 &  1.25 &  1.01 &  29 &  29 & 146 &  9.6 &  35 & + &\mbox{NF} & 256 \times 512 \times 256 \\ 
          \hline
          D1 &  5 &    36 & 71.7 & 2.00 & 0.25 &  1.25 &  1.01 &  30 &  60 &   7 &  9.3 & 4.2 & - &\mbox{PC} & 128 \times 256 \times 128 \\ 
          D2 &  5 &    36 & 71.7 & 2.00 & 0.50 &  1.25 &  1.01 &  29 &  59 &  14 &  9.5 &  60 & - &\mbox{PC} & 128 \times 256 \times 128 \\ 
          D3 &  5 &    36 & 71.7 & 2.00 & 1.00 &  1.25 &  1.01 &  29 &  59 &  29 &  9.4 &  61 & - &\mbox{PC} & 128 \times 256 \times 128 \\ 
          D4 &  5 &    36 & 71.7 & 2.00 & 2.00 &  1.25 &  1.01 &  29 &  59 &  59 &  9.5 &  43 & + &\mbox{PC} & 128 \times 256 \times 128 \\ 
          D5 &  5 &    36 & 71.7 & 2.00 & 3.33 &  1.25 &  1.01 &  29 &  59 &  98 &  9.5 &  25 & + &\mbox{NF} & 256 \times 512 \times 256 \\ 
          \hline
          E1 &  3 &    10 & 71.7 & 1.00 & 0.50 &  0.45 &  1.01 &  33 &  33 &  16 &  5.0 &  12 & - &\mbox{PC} & 144 \times 288 \times 144 \\ 
          E2 &  3 &    10 & 71.7 & 1.00 & 1.00 &  0.45 &  1.01 &  34 &  34 &  34 &  4.9 &  55 & - &\mbox{PC} & 144 \times 288 \times 144 \\ 
          E3 &  3 &    10 & 71.7 & 1.00 & 2.00 &  0.45 &  1.01 &  32 &  32 &  65 &  5.2 &  70 & + &\mbox{NF} & 144 \times 288 \times 144 \\ 
          E4 &  3 &    10 & 71.7 & 1.00 & 4.00 &  0.45 &  1.01 &  33 &  33 & 134 &  5.0 &  19 & + &\mbox{NF} & 288 \times 576 \times 288 \\ 
          \hline
          F1 &  1 &    10 & 71.7 & 1.00 & 0.50 &  0.05 &  1.01 &  40 &  40 &  20 &  1.4 &  20 & - &\mbox{PC} & 144 \times 288 \times 144 \\ 
          F2 &  1 &    10 & 71.7 & 1.00 & 1.00 &  0.05 &  1.01 &  39 &  39 &  39 &  1.4 &  43 & - &\mbox{PC} & 144 \times 288 \times 144 \\ 
          F3 &  1 &    10 & 71.7 & 1.00 & 2.00 &  0.05 &  1.01 &  38 &  38 &  76 &  1.5 &  49 & + &\mbox{PC} & 144 \times 288 \times 144 \\ 
          F4 &  1 &    10 & 71.7 & 1.00 & 4.00 &  0.05 &  1.01 &  37 &  37 & 151 &  1.5 &  30 & + &\mbox{NF} & 288 \times 576 \times 288 \\ 
          \hline
          G1 &  5 &    10 & 71.7 & 1.00 & 1.00 &  1.25 &  1.01 &  29 &  29 &  29 &  9.4 & 108 & - &\mbox{PC} & 128 \times 256 \times 128 \\ 
          G2 &  5 &    73 & 35.9 & 1.00 & 1.00 &  5.00 &  0.51 &  66 &  66 &  66 &  8.5 &  47 & + &\mbox{NF} & 256 \times 512 \times 256 \\ 
          G3 &  5 &   488 & 17.9 & 1.00 & 1.00 & 20.01 &  0.25 & 134 & 134 & 134 &  8.4 &  14 & + &\mbox{NF} & 512 \times 1024 \times 512 \\ 
          \hline
          \end{array}
          $$
          \tablefoot{
  Summary of the runs. Here, $\tilde\Omega=\Omega_0/\Omega_\odot$,
  where $\Omega_\odot=2.7\cdot10^{-6}$~s$^{-1}$ is the solar rotation
  rate,
  $\Delta t$ (in years) is the length of the time series considered, and
  the columns SSD and BCs indicate whether a small-scale dynamo is
  present and the type of latitudinal magnetic field boundary condition, respectively.
  The last column indicates the grid
  resolution. All runs have $\mathcal{L}=3\cdot10^{-5}$ and
  $\xi=0.02$. Runs C3 and G1 are the same.}
\end{table*}

\subsection{System parameters and diagnostics quantities}

The parameters governing our models are the dimensionless luminosity
\begin{equation}
\mathcal{L} = \frac{L_0}{\rho_0 (GM_\odot)^{3/2} R_\odot^{1/2}},
\end{equation}
the normalized pressure scale height at the surface,
\begin{equation}
\xi = \frac{(\gamma-1) c_{\rm V}T_1}{GM_\odot/R_\odot},
\end{equation}
with $T_1$ being the temperature at the surface in the
initial state,
the Taylor number
\begin{equation}
\Ta=(2\Omega_0 \Delta r^2/\nu)^2,
\end{equation}
where $\Delta r=r_1-r_0=0.3\,R_\odot$, as well as the
fluid, SGS, and magnetic Prandtl numbers
\begin{equation}
\Pra=\frac{\nu}{\chi_{\rm m}},\quad \Pra_{\rm SGS}=\frac{\nu}{\chiSGSm},\quad \PrM=\frac{\nu}{\eta},
\end{equation}
where $\chi_{\rm m}=K(r_{\rm m})/c_{\rm P} \rho_{\rm m}$ is the
thermal diffusivity and $\rho_{\rm m}$ is the density, both evaluated
at $r=r_{\rm m}$. We keep $\Pra=71.7$ fixed and vary
$\Pra_{\rm SGS}$ and $\PrM$ in most of our models, with the exception
of Set~G
where $\Pra_{\rm SGS}$ and $\PrM$ are set to unity and $\Pra$ is varied
by changing the value of $\nu$ (see \Table{tab:runs}).
The Rayleigh number is defined as
\begin{equation}
\Ra\!=\!\frac{GM_\odot(\Delta r)^4}{\nu \chiSGSm R_\odot^2} \bigg(-\frac{1}{c_{\rm P}}\frac{{\rm d} s_{\rm hs}}{{\rm d}r} \bigg)_{r_{\rm m}},
\label{equ:Co}
\end{equation}
where $s_{\rm hs}$ is the entropy in the hydrostatic (hs), non-convecting
state obtained from a one-dimensional model where no convection can
develop with the prescriptions of $K$ and $\chiSGS$ given above.

\begin{figure*}
  \includegraphics[width=0.5\columnwidth]{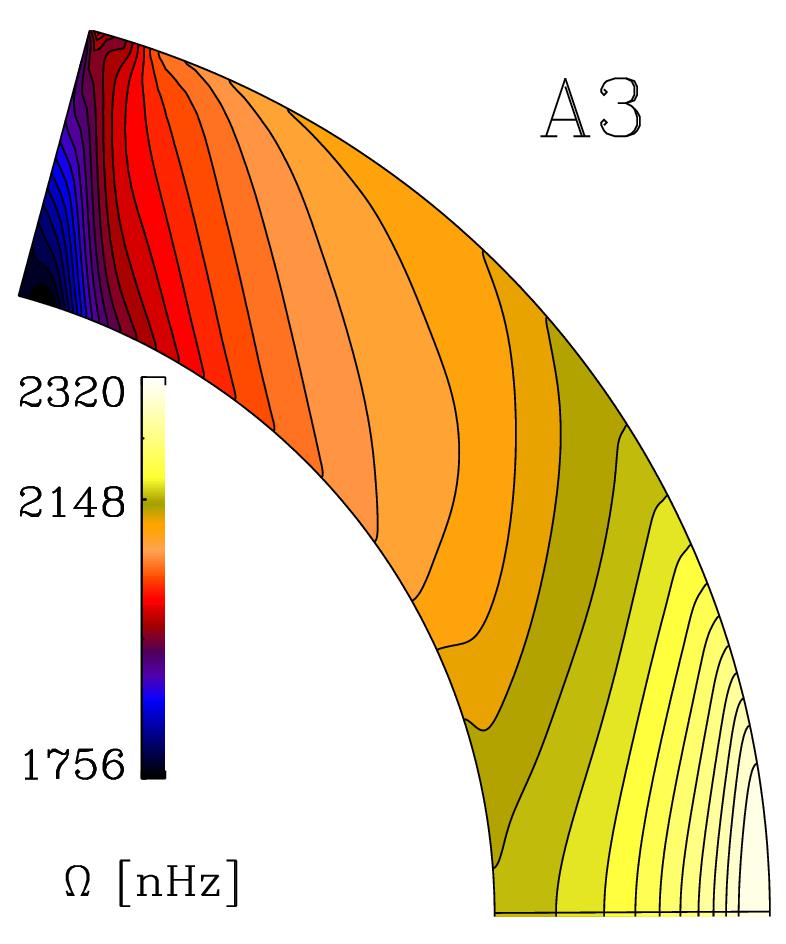}\includegraphics[width=0.5\columnwidth]{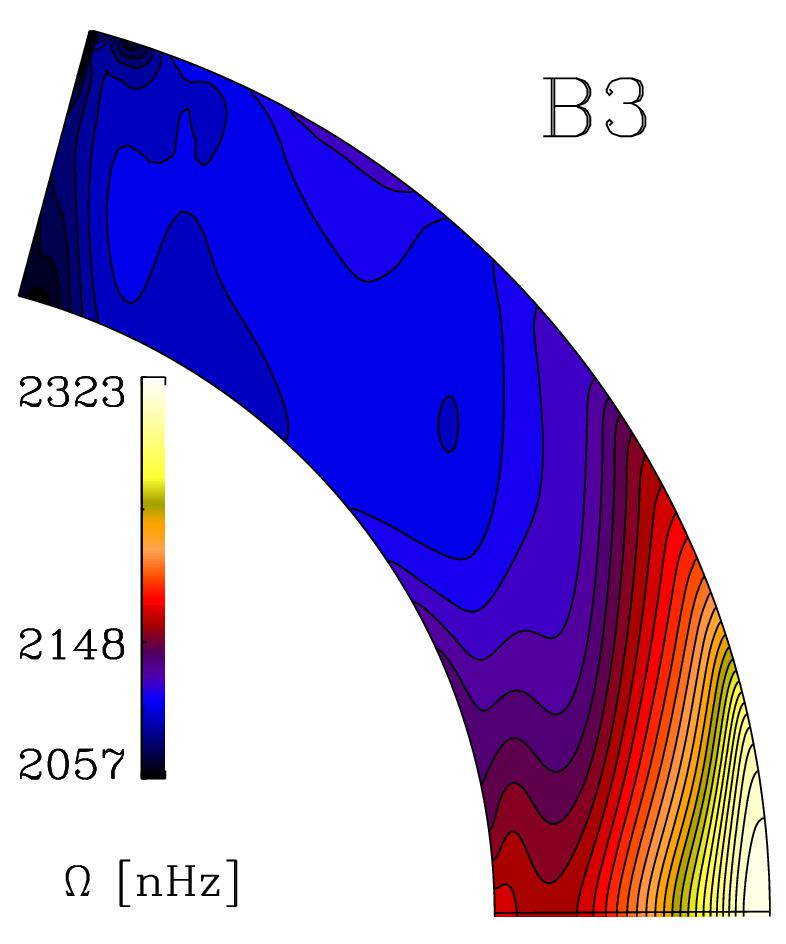}\includegraphics[width=0.5\columnwidth]{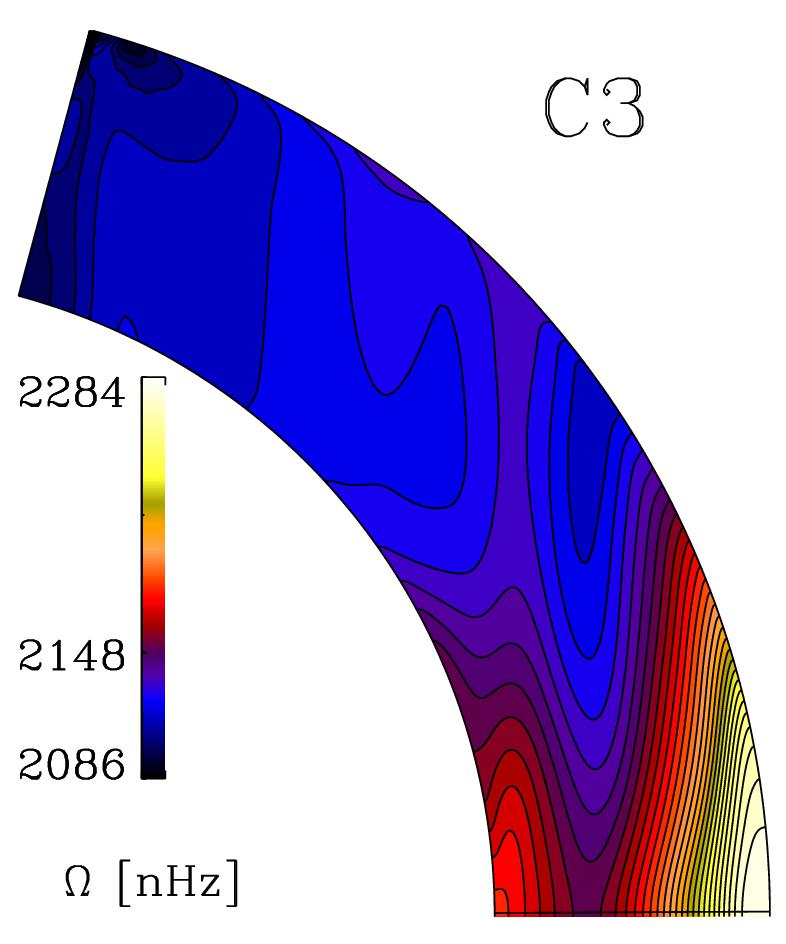}\includegraphics[width=0.5\columnwidth]{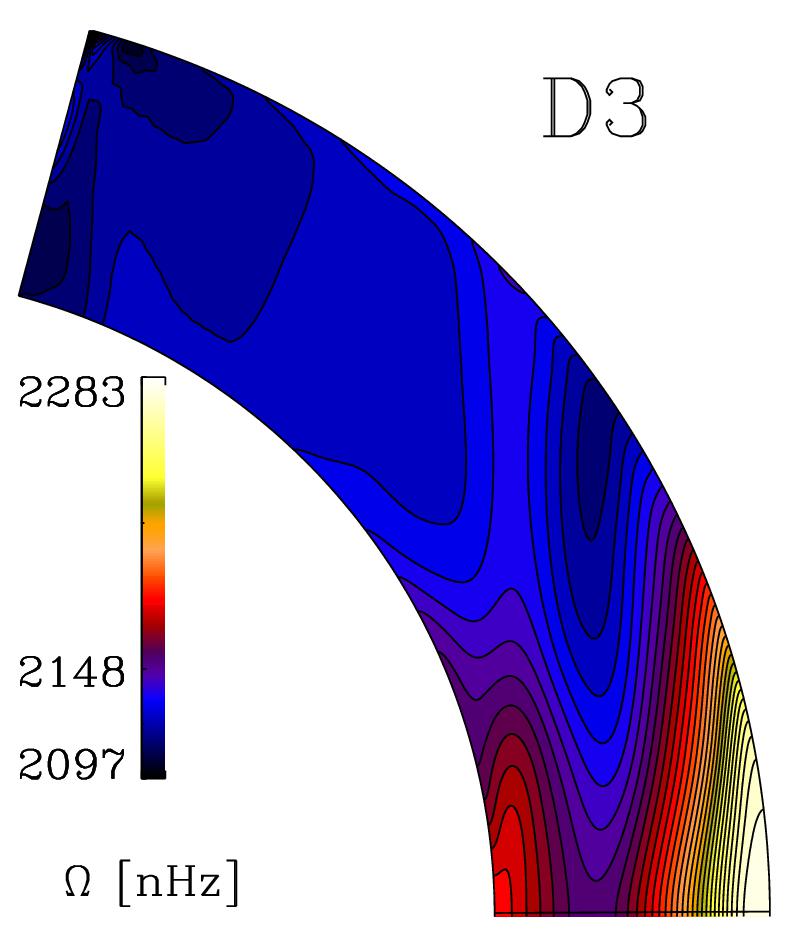}
  \includegraphics[width=0.5\columnwidth]{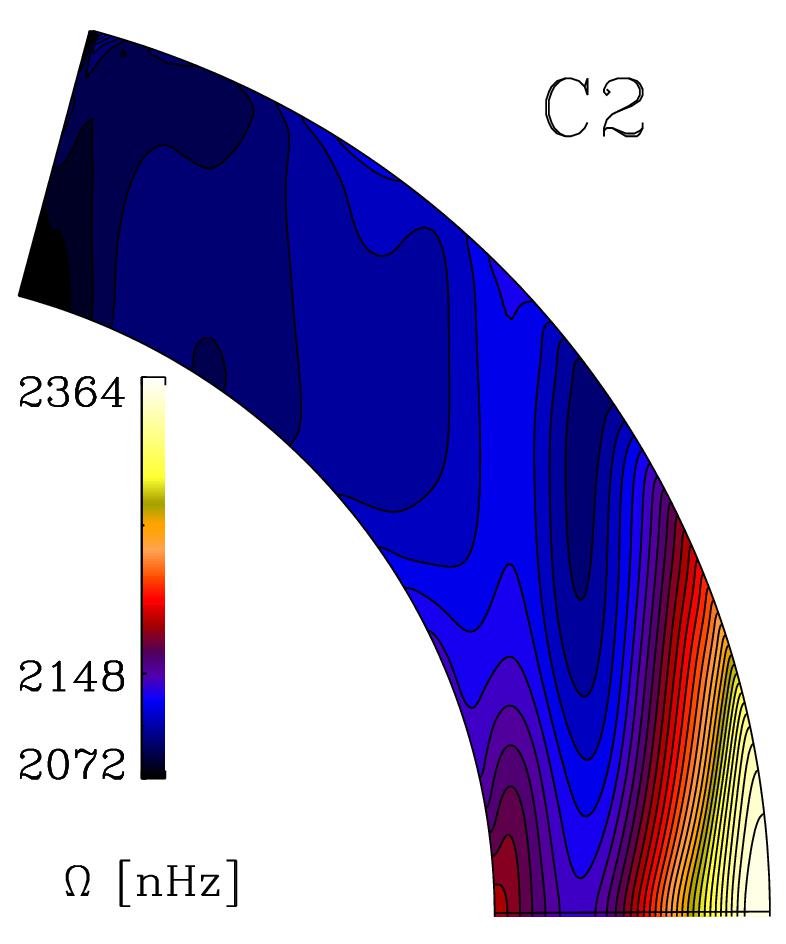}\includegraphics[width=0.5\columnwidth]{pOm_Om3_Pr100_Pm100}\includegraphics[width=0.5\columnwidth]{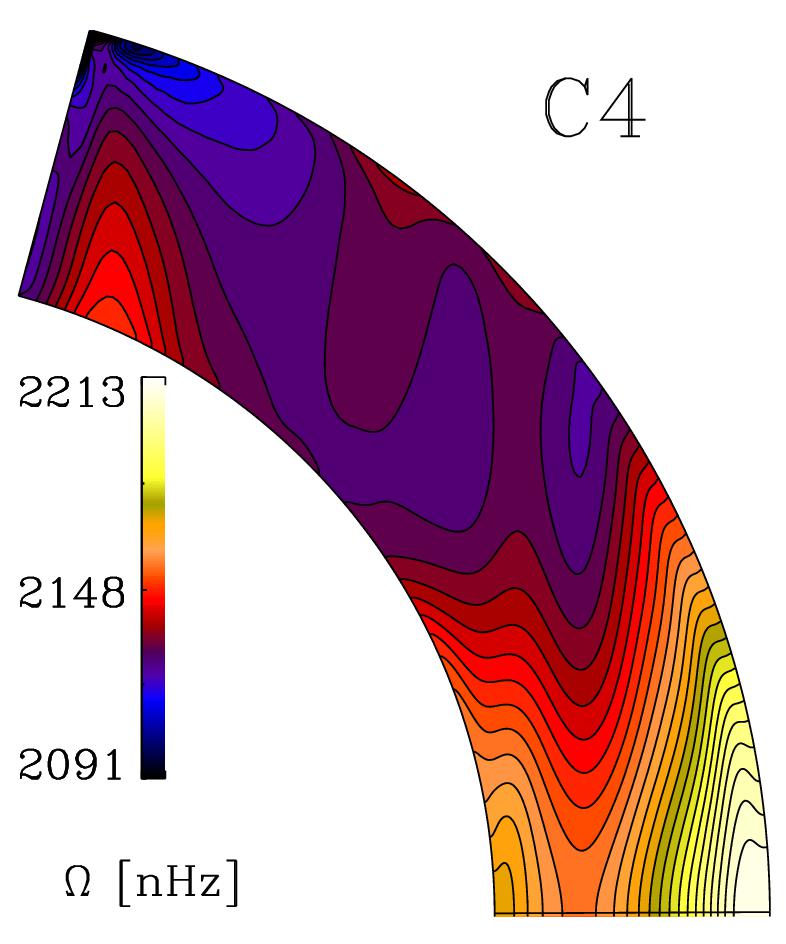}\includegraphics[width=0.5\columnwidth]{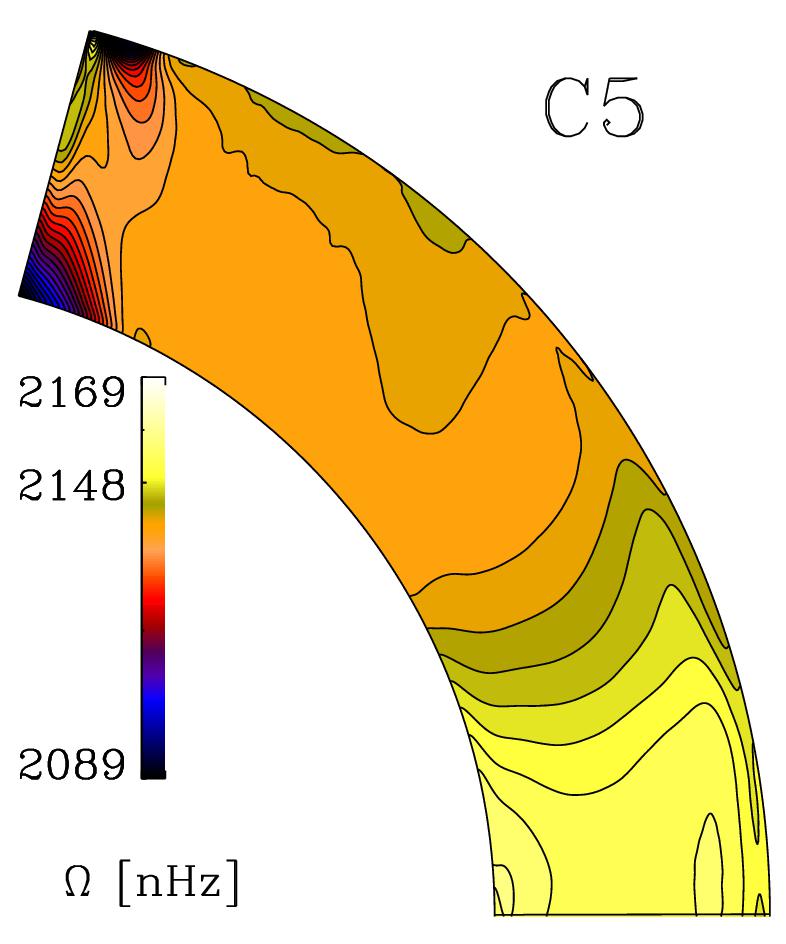}
\caption{{\it Top row:} temporally averaged rotation profiles from
  Runs~A3, B3, C3, and D3 with $\PrM=1$ and $\PraSGS$ varying from 0.25 (left) to 2
  (right). {\it Lower row:} the same as above but for Runs~C2, C3, C4,
  and C5 with $\PraSGS=1$ and $\PrM$ varying from 0.5 (left) to 5 (right).}
\label{fig:pOm}
\end{figure*}

The remaining parameters are used only as diagnostics. These include
the fluid and magnetic Reynolds numbers, and the P\'eclet number
\begin{equation}
\Rey=\frac{\urms}{\nu \kf},\quad \ReM=\frac{\urms}{\eta \kf},\quad
\Pe=\frac{\urms}{\chiSGSm \kf},
\end{equation}
where $\kf=2\pi/\Delta r\approx21 R_\odot^{-1}$ is an estimate of the
wavenumber of the largest eddies. Rotational influence on the flow is
given by the Coriolis number
\begin{equation}
\Co=\frac{2\Omega_0}{\urms \kf},
\label{eq:Coriolis}
\end{equation}
where $\urms=\sqrt{(3/2)\brac{U_r^2+U_\theta^2}_{r\theta\phi t}}$ is
the rms velocity and the subscripts indicate averaging over $r$,
$\theta$, $\phi$, and a time interval during which the run is
thermally relaxed. We omit the contribution from the azimuthal
velocity in $\urms$, because it is dominated by differential
rotation \citep{KMGBC11}.

We define mean quantities as averages over the $\phi$-coordinate and
denote them by overbars. We also often average the data in time over
the period of the simulations where thermal energy, differential
rotation, and large-scale magnetic fields have reached statistically
saturated states.

The simulations are performed with the {\sc Pencil
  Code}\footnote{http://pencil-code.github.com/}, which uses a
high-order finite difference method for solving the compressible
equations of magnetohydrodynamics.

\section{Data analysis: $D^2$ statistic}
\label{DataAnalysis}

To detect possible cycles and to estimate their average lengths
we have chosen to use $D^2$ phase dispersion
statistic \citep{Pelt83}. It has recently been applied to irregularly
spaced long-term photometry of solar-like stars
\citep{LMOPHHJS13,Olspert15} as well as to more regularly sampled
magnetoconvection simulation data \citep{KKKBOP15,KKOBWKP16}. In the
previous applications the statistic has been used exclusively on
one-dimensional time series (e.g.\ by fixing a certain latitude and radius
in the azimuthally averaged data).
In the current study we use a
generalized form of the statistic given by
\begin{equation}\label{eq:D2}
D^2(P,\Delta t_{\rm coh})\!=\!\frac{\sum\limits_{i=1}^{N-1} {\sum\limits_{j=i+1}^N {g(t_i}} ,t_j ,P,\Delta t_{\rm coh})||{\bm f}(t_i ) - {\bm f}(t_j )||^2}{2\sigma^2\sum\limits_{i = 1}^{N - 1} {\sum\limits_{j = i + 1}^N {g(t_i } } ,t_j ,P,\Delta t_{\rm coh})},
\end{equation}
where ${\bm f}(t_i)$ is the vector of observed variables at time
moment $t_i$, $\sigma^2=N^{-2}\sum_{i,j>i}||{\bm f}(t_i ) -
{\bm f}(t_j )||^2$ is the variance of the full time series,
$g(t_i,t_j,P,\Delta t_{\rm coh})$ is the selection function, which is
significantly greater than zero only when
\begin{eqnarray}
  t_j  - t_i  &\approx& kP,k =  \pm 1, \pm 2, \ldots, \quad \mbox{and}\\
\left| {t_j  - t_i } \right| &\lessapprox& \Delta t_{\rm coh} = l_{\rm coh}P,
\end{eqnarray}
where $P$ is the trial period and $\Delta t_{\rm coh}$ is the so-called
coherence time, which is the measure of the width of the sliding time window
wherein the data points are taken into account by the statistic.
The number of trial periods fitting into this interval, $l_{\rm coh}=\Delta t_{\rm coh}/P$, is called a coherence length.
With this definition there is no restriction on the dimensionality of
the data, but Eq.~(\ref{eq:D2}) leaves open the choice of the vector norm.
In most cases it is natural to use the Euclidean norm,
which we also do in our analysis.
We use this statistic to analyze the radial and azimuthal components of the magnetic field
at regions near the surface over latitude intervals, where the cycles are the most
pronounced (see Sect.~\ref{D2res}).

\begin{figure*}
\centering
\includegraphics[width=0.8\textwidth]{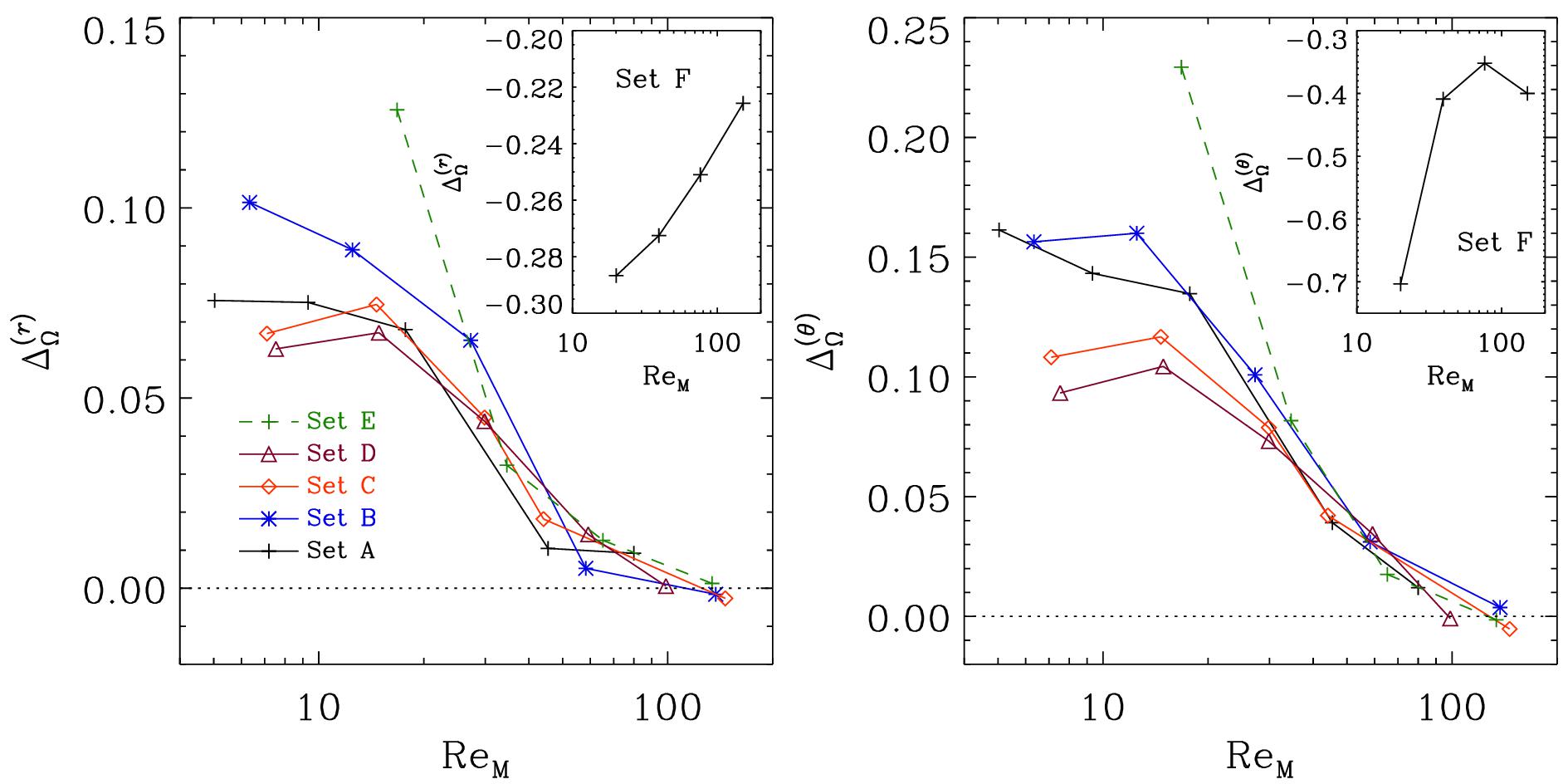}
\caption{Estimates of radial and latitudinal differential rotation
  $\Delta_\Omega^{(r)}$ and $\Delta_\Omega^{(\theta)}$, respectively,
  according to Eq.~(\ref{equ:pDRt}) for Sets~A--F as indicated by the
  legend.}
\label{fig:pdrot}
\end{figure*}

\section{Results} \label{sect:results}

We perform six sets of simulations (Sets~A--F), each with a
constant value of
$\PraSGS$ but changing $\PrM$ in the range $0.25\leq\PrM\leq5$; see
Table~\ref{tab:runs}. Furthermore, in an additional Set~G we fix
$\PraSGS=\PrM=1$ and
vary the Reynolds and P\'eclet numbers. The rotation rate is varied such
that Sets~A--D and G have $\Omega_0=5\Omega_\odot$, whereas in Sets~E
and F we use
$3\Omega_\odot$ and $\Omega_\odot$, respectively. Sets E and F are
included in order to study the robustness of our findings at slower
rotation. In some sets (A, D,
E, and F) numerical problems prevented the use of $\PrM=5$. In those
cases we used a lower value that produces
numerically stable solutions.
The latitudinal PC boundary conditions lead to numerical problems due to
an unidentified instability at high magnetic Reynolds numbers near the
latitudinal boundaries. Cases
where this occurred have been rerun with the NF conditions. We do not
find major qualitative differences in the behavior of the large-scale
field between PC and NF runs with otherwise identical
parameters. Furthermore, we omit the $\PrM=0.25$ runs that do not lead
to dynamos in Sets~E and F.
We note that Run~B2 has been presented as Run~II in \cite{WKKB14},
Run~D3 as Run~A1 in \cite{WKKB16,WRKKB16}, and Run~C3 covers the first
120~years of
the run presented in \cite{KKOBWKP16}.
Furthermore, runs similar to Run~D3 (but with $\PrM=2.5$ instead of $2.0$)
have been presented as Runs~B4m and C1 in \cite{KMB12,KMCWB13}.

\subsection{Large-scale flows and their generators}

\subsubsection{Differential rotation and meridional circulation}

The sign of the radial gradient of $\Omega$ plays a crucial role in
determining the propagation direction of dynamo waves in
$\alpha\Omega$ dynamos: with a positive $\alpha$ effect in the
northern hemisphere, a negative radial gradient of $\Omega$ is
required to obtain solar-like equatorward migration and vice versa
\citep{Pa55a,Yo75}. It is remarkable that this rule also seems to
apply to the fully nonlinear convective dynamo simulations
\citep{WKKB14,WRKKB16}. The current simulations can produce equatorward
migration only in cases where a region with a negative radial gradient
of $\Omega$ occurs at mid-latitudes \citep[e.g.][]{KMB12,KMCWB13,ABMT15}
or if the sign of the kinetic helicity, which is a proxy of the $\alpha$
effect, is inverted in the bulk of the
convection zone \citep{DWBG16}.

The simulations of \cite{KMB12}, 
\cite{ABMT15}, \cite{KKOBWKP16}, and \cite{WKKB16} showing equatorward
migration and a region of negative radial shear at mid-latitudes had
$\PraSGS\gtrsim1$. This is in contrast to earlier simulations
with $\PraSGS<1$, which did not show
equatorward
migration \citep[e.g.][]{BMT04,BMBBT11,NBBMT13} and had consistently
positive gradients of $\Omega$.
Given that the dynamo wave propagation is apparently heavily influenced by
this, it is important to study the effect that $\PraSGS$ has on the
rotation profiles.

We show representative results of the temporally averaged
rotation profiles as a function of $\PraSGS$ from
Runs~A3, B3, C3, and D3 in the top row of Fig.~\ref{fig:pOm}. In the
lowest SGS Prandtl number case (Run~A3), the
angular velocity $\Omega=\Omega_0 + \mean{U}_\phi/r\sin\theta$
decreases monotonically from the equator toward the
poles. Much of the latitudinal variation occurs at high latitudes near
the latitudinal boundaries. However, the rotation profile is
qualitatively similar to those obtained from low $\PraSGS$ models
in fully spherical shells \citep[e.g.][]{BMT04,BBBMT10}.
In Run~B3, a dip at mid-latitudes is developing,
which is seen to deepen in the higher $\PraSGS$ runs C3 and D3. We
note that a similar
transition occurs also when the density stratification is increased
with $\PraSGS=2\ldots5$ \citep{KMB11,KMCWB13}. The overall magnitude
of the differential rotation also
decreases from low to high SGS Prandtl numbers. However, much of this
variation occurs already between Runs~A3 and B3 whereas the
differences between Runs~B3, C3, and D3 are much smaller.

The time-averaged rotation profiles from runs with $\PraSGS=1$ with varying $\PrM$
are shown in the lower row of \Fig{fig:pOm} from Runs~C2--C5. The
absolute shear decreases steeply as $\PrM$ and $\ReM$ increase so that
in the highest $\ReM$ case the differential rotation is appreciable
only near the latitudinal boundaries. There are also qualitative
changes such that the negative shear layer at mid-latitudes is almost
absent in Run~C5 and a near-surface shear layer is developing in
Runs~C4 and C5.

To quantify the radial and latitudinal differential rotation we use
the quantities \citep{KMCWB13}
\begin{eqnarray}
\Delta_\Omega^{(r)}=\frac{\Omega_{\rm eq}-\Omega_{\rm bot}}{\Omega_{\rm eq}},
\quad
\Delta_\Omega^{(\theta)}=\frac{\Omega_{\rm eq}-\Omega_{\rm pole}}{\Omega_{\rm eq}},\label{equ:pDRt}
\end{eqnarray}
where $\Omega_{\rm eq}=\Omega(r_1,\pi/2)$ and $\Omega_{\rm
  bot}=\Omega(r_0,\pi/2)$ are the rotation rates at the surface
and at the bottom of the convection zone at the equator. Furthermore,
$\Omega_{\rm
  pole}=\onehalf[\Omega(r_1,\theta_0)+\Omega(r_1,\pi-\theta_0)]$
is the average rotation rate between the latitudinal boundaries at the
outer boundary. We show $\Delta_\Omega^{(r)}$ and
$\Delta_\Omega^{(\theta)}$ for the runs of Sets~A--F in
Fig.~\ref{fig:pdrot}. We find that both radial and latitudinal
differential rotation are modestly quenched for $\ReM\lesssim30$ in
Sets~A--E. For higher values of $\ReM$, both $\Delta_\Omega^{(r)}$ and
$\Delta_\Omega^{(\theta)}$ decrease steeply, and for the highest values
of $\ReM$ the differential rotation is almost completely
quenched. Set~F with the lowest rotation rate stands apart from the
other runs. The major difference in this set is that the differential
rotation is anti-solar, i.e.\ with a slow equator and faster
poles. There the radial differential rotation decreases monotonically,
but the decrease is only roughly 20 per cent in the range
$\ReM=20\ldots151$ whereas $\Delta_\Omega^{(\theta)}$ remains roughly
constant above $\ReM=39$. A possible explanation to the difference
between Set~F and the other sets is that the mean magnetic fields are
stronger in the latter sets (see \Table{tab:energies} and
Sect.~\ref{sec:satu}) leading to a
stronger backreaction to the flow.

\begin{table*}[t!]
\centering
\caption[]{Volume and time-averaged kinetic and magnetic energy
  densities realized in the
  simulations in units of $10^5$~J~m$^{-3}$.}
\label{tab:energies}
      $$
          \begin{array}{p{0.025\linewidth}ccccccccccccccccl}
            \hline
            \hline
            \noalign{\smallskip}
            Run & E_{\rm kin} & \delta E_{\rm kin} & E_{\rm kin}^{\rm DR} & \delta E_{\rm kin}^{\rm DR} & E_{\rm kin}^{\rm MC} & \delta E_{\rm kin}^{\rm MC} & E_{\rm kin}^{\rm fluct} & \delta E_{\rm kin}^{\rm fluct} & E_{\rm mag} & \delta E_{\rm mag} & E_{\rm mag}^{\rm tor} & \delta E_{\rm mag}^{\rm tor} & E_{\rm mag}^{\rm pol} & \delta E_{\rm mag}^{\rm pol} & E_{\rm mag}^{\rm fluct} & \delta E_{\rm mag}^{\rm fluct}\\
            \hline
            A1 & 12.981 &  0.281 & 12.107 &  0.219 &  0.011 &  0.002 &  0.863 &  0.060 &  0.000 &  0.000 &  0.000 &  0.000 &  0.000 &  0.000 &  0.000 &  0.000 \\ 
            A2 & 12.485 &  0.307 & 11.741 &  0.207 &  0.012 &  0.003 &  0.732 &  0.183 &  0.000 &  0.000 &  0.000 &  0.000 &  0.000 &  0.000 &  0.000 &  0.000 \\ 
            A3 &  8.067 &  1.150 &  7.362 &  1.133 &  0.009 &  0.000 &  0.695 &  0.017 &  0.293 &  0.023 &  0.197 &  0.019 &  0.011 &  0.002 &  0.086 &  0.011 \\ 
            A4 &  1.206 &  0.117 &  0.434 &  0.132 &  0.007 &  0.001 &  0.766 &  0.015 &  0.834 &  0.130 &  0.367 &  0.077 &  0.041 &  0.006 &  0.426 &  0.051 \\ 
            A5 &  1.088 &  0.059 &  0.282 &  0.121 &  0.006 &  0.001 &  0.800 &  0.081 &  1.116 &  0.272 &  0.398 &  0.200 &  0.070 &  0.049 &  0.649 &  0.141 \\ 
            \hline
            B1 & 24.510 &  0.144 & 23.087 &  0.144 &  0.017 &  0.001 &  1.406 &  0.068 &  0.000 &  0.000 &  0.000 &  0.000 &  0.000 &  0.000 &  0.000 &  0.000 \\ 
            B2 & 20.505 &  1.056 & 19.036 &  1.028 &  0.015 &  0.001 &  1.454 &  0.033 &  0.680 &  0.021 &  0.350 &  0.021 &  0.041 &  0.003 &  0.289 &  0.007 \\ 
            B3 &  7.233 &  1.775 &  5.722 &  1.674 &  0.014 &  0.001 &  1.496 &  0.128 &  0.982 &  0.139 &  0.431 &  0.105 &  0.051 &  0.014 &  0.500 &  0.055 \\ 
            B4 &  1.816 &  0.081 &  0.558 &  0.076 &  0.009 &  0.000 &  1.248 &  0.045 &  1.463 &  0.072 &  0.387 &  0.081 &  0.089 &  0.007 &  0.987 &  0.016 \\ 
            B5 &  1.169 &  0.139 &  0.167 &  0.066 &  0.007 &  0.001 &  0.995 &  0.074 &  2.325 &  1.023 &  0.501 &  0.487 &  0.106 &  0.040 &  1.718 &  0.496 \\ 
            \hline
            C1 &  7.248 &  0.364 &  5.695 &  0.359 &  0.014 &  0.000 &  1.538 &  0.011 &  0.000 &  0.000 &  0.000 &  0.000 &  0.000 &  0.000 &  0.000 &  0.000 \\ 
            C2 &  8.484 &  0.635 &  6.705 &  0.614 &  0.018 &  0.000 &  1.760 &  0.021 &  0.637 &  0.057 &  0.145 &  0.018 &  0.118 &  0.013 &  0.374 &  0.026 \\ 
            C3 &  4.299 &  0.258 &  2.641 &  0.220 &  0.015 &  0.001 &  1.642 &  0.038 &  1.035 &  0.111 &  0.258 &  0.044 &  0.102 &  0.014 &  0.676 &  0.056 \\ 
            C4 &  2.630 &  0.382 &  1.069 &  0.323 &  0.012 &  0.001 &  1.549 &  0.058 &  1.308 &  0.144 &  0.357 &  0.076 &  0.078 &  0.011 &  0.873 &  0.087 \\ 
            C5 &  1.310 &  0.062 &  0.156 &  0.041 &  0.007 &  0.000 &  1.146 &  0.063 &  2.138 &  0.071 &  0.309 &  0.074 &  0.089 &  0.008 &  1.740 &  0.121 \\ 
            \hline
            D1 &  6.895 &  0.149 &  5.211 &  0.119 &  0.015 &  0.001 &  1.668 &  0.032 &  0.000 &  0.000 &  0.000 &  0.000 &  0.000 &  0.000 &  0.000 &  0.000 \\ 
            D2 &  6.775 &  0.055 &  4.977 &  0.049 &  0.017 &  0.000 &  1.781 &  0.008 &  0.653 &  0.013 &  0.159 &  0.007 &  0.133 &  0.003 &  0.361 &  0.006 \\ 
            D3 &  3.978 &  0.426 &  2.296 &  0.352 &  0.014 &  0.000 &  1.668 &  0.074 &  0.873 &  0.051 &  0.171 &  0.036 &  0.098 &  0.012 &  0.604 &  0.024 \\ 
            D4 &  2.012 &  0.130 &  0.545 &  0.110 &  0.010 &  0.000 &  1.457 &  0.020 &  0.987 &  0.041 &  0.145 &  0.039 &  0.074 &  0.003 &  0.768 &  0.026 \\ 
            D5 &  1.545 &  0.044 &  0.313 &  0.065 &  0.008 &  0.000 &  1.224 &  0.032 &  1.607 &  0.139 &  0.261 &  0.082 &  0.068 &  0.006 &  1.278 &  0.052 \\ 
            \hline
            E1 & 26.246 &  0.545 & 23.956 &  0.571 &  0.035 &  0.001 &  2.255 &  0.066 &  0.366 &  0.013 &  0.107 &  0.008 &  0.027 &  0.002 &  0.232 &  0.017 \\ 
            E2 &  3.904 &  1.355 &  1.881 &  1.289 &  0.022 &  0.002 &  2.000 &  0.064 &  1.252 &  0.137 &  0.424 &  0.102 &  0.070 &  0.011 &  0.758 &  0.064 \\ 
            E3 &  2.236 &  0.336 &  0.542 &  0.265 &  0.017 &  0.001 &  1.677 &  0.069 &  1.377 &  0.142 &  0.235 &  0.064 &  0.074 &  0.006 &  1.068 &  0.087 \\ 
            E4 &  2.170 &  0.114 &  0.428 &  0.151 &  0.017 &  0.000 &  1.724 &  0.056 &  1.700 &  0.118 &  0.267 &  0.068 &  0.061 &  0.007 &  1.372 &  0.075 \\ 
            \hline
            F1 &  7.383 &  2.069 &  4.140 &  2.086 &  0.119 &  0.006 &  3.124 &  0.141 &  0.836 &  0.106 &  0.154 &  0.038 &  0.101 &  0.039 &  0.581 &  0.095 \\ 
            F2 &  4.845 &  0.094 &  1.524 &  0.147 &  0.099 &  0.002 &  3.222 &  0.055 &  0.871 &  0.055 &  0.083 &  0.002 &  0.063 &  0.017 &  0.724 &  0.038 \\ 
            F3 &  4.432 &  0.111 &  1.307 &  0.028 &  0.086 &  0.002 &  3.039 &  0.093 &  0.910 &  0.088 &  0.067 &  0.009 &  0.042 &  0.012 &  0.801 &  0.066 \\ 
            F4 &  4.015 &  0.095 &  1.057 &  0.092 &  0.079 &  0.001 &  2.880 &  0.016 &  1.177 &  0.050 &  0.061 &  0.004 &  0.036 &  0.004 &  1.079 &  0.041 \\ 
            \hline
            G1 &  4.299 &  0.258 &  2.641 &  0.220 &  0.015 &  0.001 &  1.642 &  0.038 &  1.035 &  0.111 &  0.258 &  0.044 &  0.102 &  0.014 &  0.676 &  0.056 \\ 
            G2 &  2.782 &  0.947 &  1.026 &  0.680 &  0.014 &  0.003 &  1.742 &  0.264 &  1.501 &  0.233 &  0.288 &  0.086 &  0.090 &  0.008 &  1.123 &  0.161 \\ 
            G3 &  2.401 &  0.608 &  0.654 &  0.487 &  0.014 &  0.002 &  1.733 &  0.224 &  1.938 &  0.212 &  0.372 &  0.122 &  0.087 &  0.034 &  1.479 &  0.124 \\ 
            \hline
          \end{array}
          $$
          \tablefoot{Runs~A1, A2, B1, C1, and D1 do not have
            dynamos. The $\delta$-quantities represent the
            variations in time. These are computed by first dividing the
            time series in three equally long parts and temporally
            averaging over each of these. Then the greatest deviation
            of these quantities from the average over the whole time
            series is taken to represent the variations in time.}
\end{table*}

\begin{figure}
\includegraphics[width=0.97\columnwidth]{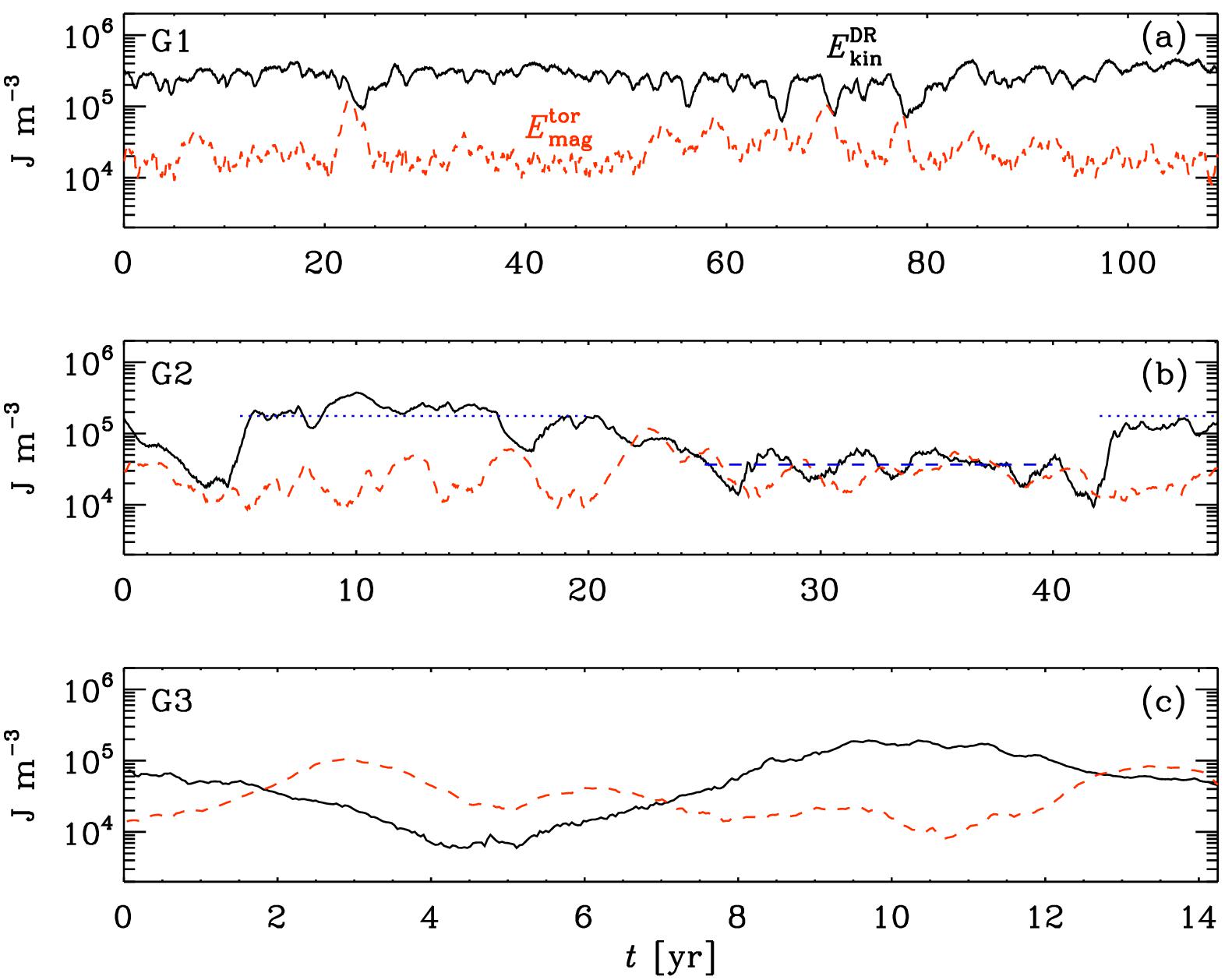}
\includegraphics[width=0.97\columnwidth]{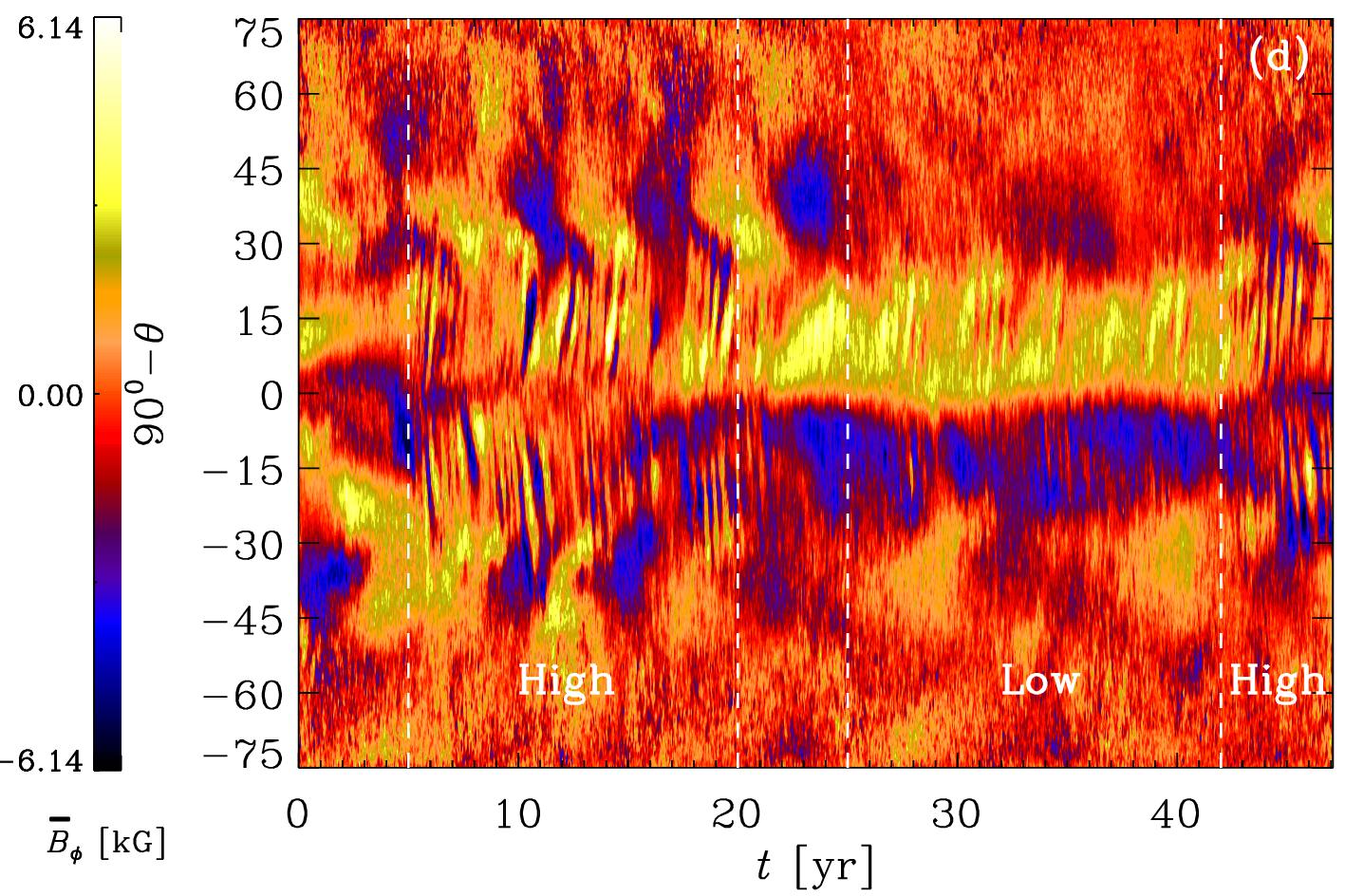}
\includegraphics[width=0.97\columnwidth]{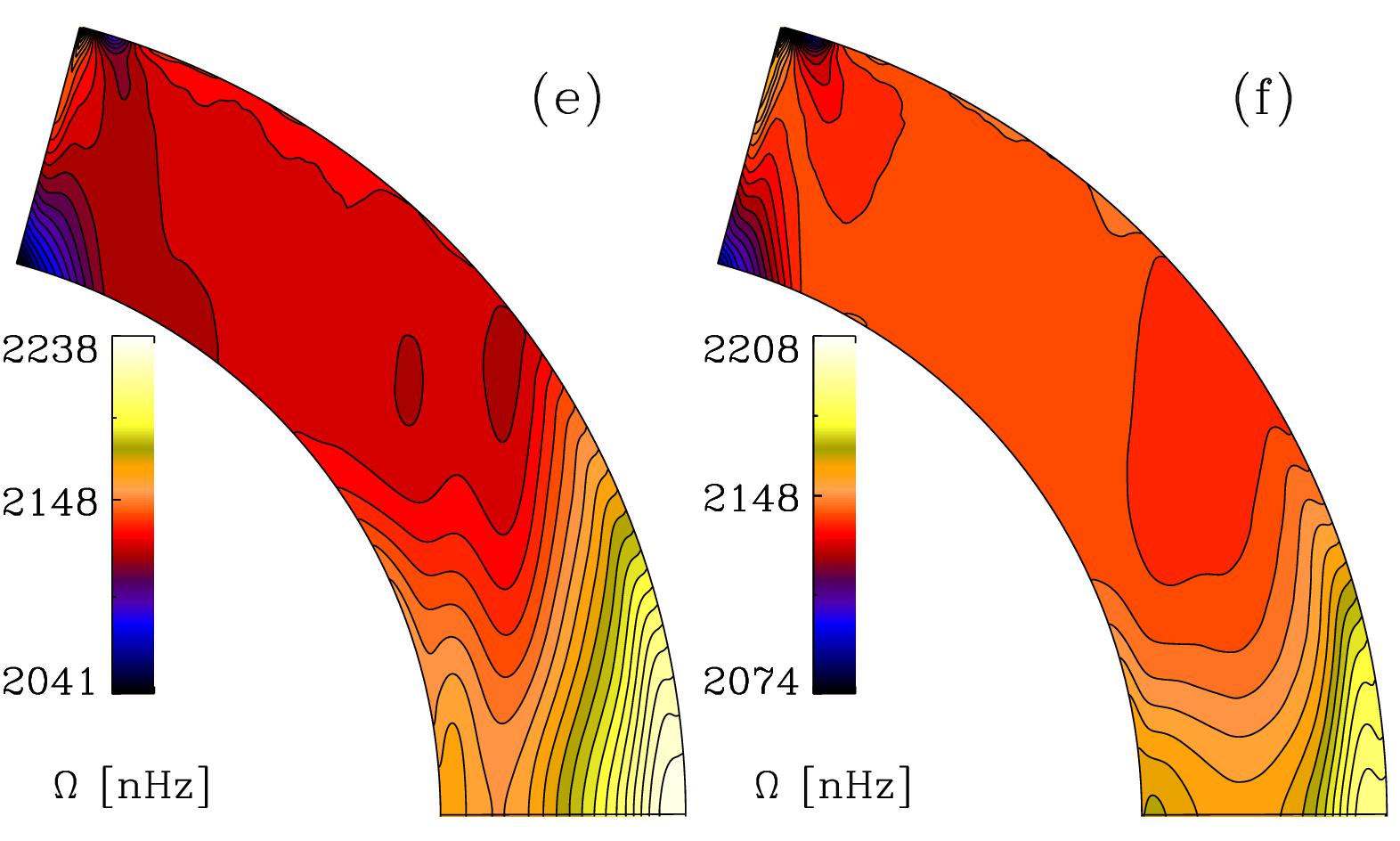}
\vspace{-0.2cm}
\caption{Energies of differential rotation (black solid lines) and
  mean toroidal magnetic field (red dashed) as functions of time from
  Runs~G1, G2, and G3 (panels (a)--(c)). Panel (d) shows the
  azimuthally averaged azimuthal magnetic field at $r=0.98R_\odot$
  from Run~G2. The vertical dashed lines indicate the `High' and `Low'
  states of differential rotation. Panels (e) and (f) show the time
  averaged rotation profiles in Run~G2 from the high and low states
  indicated as blue dotted and dashed lines in panel (b).}
\label{fig:penet}
\end{figure}

We list the kinetic energy densities of the total flow $E_{\rm kin}=\brac{\onehalf
\mean{\rho \UUU^2}}_V$, differential rotation $E_{\rm kin}^{\rm
  DR}=\brac{\onehalf \mean{\rho} \mean{U}_\phi^2}_V$, meridional circulation
$E_{\rm kin}^{\rm MC}=\brac{\onehalf \mean{\rho}
(\mean{U}_r^2+\mean{U}_\theta^2)}_V$, the fluctuating velocity
$E_{\rm kin}^{\rm fluct}=\brac{\onehalf \mean{\rho \uuu^2}}_V$ and the
magnetic energy densities related to the total field $E_{\rm
  mag}=\brac{\mean{\BBB^2}/2\mu_0}_V$, azimuthally averaged toroidal
$E_{\rm mag}^{\rm tor}=\brac{\mean{B}_\phi^2/2\mu_0}_V$ and poloidal
fields $E_{\rm mag}^{\rm
  pol}=\brac{(\mean{B}_r^2+\mean{B}_\theta^2)/2\mu_0}_V$, and the
fluctuating magnetic energy $E_{\rm mag}^{\rm
  fluct}=\brac{\mean{\bbb^2}/2\mu_0}_V$ with $\uuu=\UUU-\meanv{U}$,
$\bbb=\BBB-\meanv{B}$ and where $\brac{\ }_V$ indicates a volume
average, in our simulations
in Table~\ref{tab:energies}. We find that the kinetic energy decreases
monotonically as the magnetic Reynolds number is increased
irrespective of the SGS Prandtl number. This is mostly due to
quenching of the differential rotation whereas the fluctuating kinetic
energy is much less affected. Differences between the sets of runs are
large, however. In Set~F, the total kinetic energy drops by less than a
factor of two and the energy of the differential rotation by a factor
of four, whereas in Sets~A, B, C, and E, the $E_{\rm kin}$ reduces by
roughly an order of magnitude and $E_{\rm kin}^{\rm (DR)}$ by two
orders of magnitude. Set~D falls between the two extreme cases. The
energy of the meridional flow is
negligible in comparison to both differential rotation and fluctuating
(non-axisymmetric) contributions.

We note, however, that the temporal variations of differential
rotation
increase as the magnetic Reynolds number is increased, see
\Table{tab:energies}. This is demonstrated in
\Fig{fig:penet}(a)--(c) where the energies of the differential rotation and
mean toroidal magnetic field are shown for Runs~G1--G3. In the
lowest-$\ReM$ case, both are fairly stable with $E_{\rm mag}^{\rm tor}$
being typically an order of magnitude smaller than $E_{\rm kin}^{\rm
  DR}$ with a few excursions with strong magnetic field and weak
differential rotation (e.g.\ around $t\approx22$~yrs and
$t\approx50-80$~yrs). In an extended version of this run, the quiescent
magnetic field leads to stronger than average differential rotation for the
last 70~years of that run \citep[see Figs.\ 4(a) and 4(b)
  of][]{KKOBWKP16}. In Run~G2 with $\ReM=66$, two more distinct states
appear to be
present: either the differential rotation is strong and the magnetic
fields weak ($t\approx5-20$ and $t\gtrsim42$ years) or the two are
comparable ($t\approx25-42$ years). These events are associated with a
change of the large-scale dynamo mode from an oscillatory equatorward
migrating mode (strong DR, weak magnetic field) to a quasi-stationary
one (DR and magnetic field energies comparable), see
\Fig{fig:penet}(d). Panels (e) and (f) of \Fig{fig:penet} show
that the latitudinal differential rotation
decreases by roughly 30 per cent from the high to the low state.
Similar but apparently more violent variations are seen in the
highest-$\ReM$ case (Run~G3), but there the time series is too short to
draw solid conclusions.

Our results appear to stand apart from similar studies in full
spherical shells \citep[e.g.][]{NBBMT13,HRY16} in that the differential
rotation is strongly quenched as a function of the magnetic Reynolds
number. However, in \cite{NBBMT13} the values of $\Rm'$ ($=2\pi\ReM$)
correspond to
a range of $8\ldots32$ in $\ReM$ in our units where the radial and
latitudinal differential rotation decrease by about 30 per cent. This
is roughly consistent with our results. On the other hand,
\cite{HRY16} reach higher values of $\ReM$ than in the present study,
but no strong quenching is reported. This can be because their models
are rotating substantially slower than ours, leading to weaker
magnetic fields and a weaker back reaction to the flow. 
Furthermore, in these models, the differential rotation is strongly
influenced by their SGS heat flux, which transports one third of the
luminosity.
Another
obvious candidate for explaining
the difference is the wedge
geometry used in the current simulations. However, we note that
earlier simulations with a similar setup did not show a marked
trend in the energy of the differential rotation as the azimuthal
extent of the domain was varied \citep[see Table 1
  of][]{KMCWB13}. However, results of Boussinesq simulations of
convective dynamos have shown a similar change as a function of the
magnetic Prandtl number \citep{SPD12}. The drop in the amplitude of
the differential rotation was associated with a change in the dynamo
mode from an oscillatory multipolar solution to a quasi-steady dipolar
configuration \citep[cf.\ Fig.~15 of][]{SPD12} that prevents strong
differential rotation from developing.
We do not find a strong dipole component in our simulations (see,
Sect.~\ref{sec:dynmod}). However, the strong suppression of the
differential rotation often coincides with the appearance of a
small-scale dynamo (see \Table{tab:runs} and the discussion in the
\Sec{sec_Amt}) or a change in the large-scale dynamo mode as
discussed above.

\subsubsection{Angular momentum transport}
\label{sec_Amt}

The azimuthally averaged $z$-component of the angular momentum is
governed by the equation:
\begin{eqnarray}
&& \frac{\pd}{\pd t} (\mean{\rho} \varpi^2 \Omega)
+ \bm\nabla\bm\cdot\{\varpi [\varpi\mean{\rho \bm{U}}\Omega
+ \mean{\rho}\ \mean{u_\phi {\bm u}}
- 2\nu\mean{\rho} \mean{\bm{\mathsf{S}}}\bm\cdot\bm{\hat\phi}
\nonumber \\ && \hspace{4cm}
-\mu_0^{-1}(\mean{B}_\phi\meanv{B} + \mean{b_\phi {\bm b}})] \}=0,
\label{equ:angmom}
\end{eqnarray}
where $\varpi=r\sin\theta$ is the lever arm and velocity and
magnetic field have been decomposed into mean and fluctuating
parts according to $U_i = \mean{U}_i + u_i$ and $B_i = \mean{B}_i +
b_i$. We will denote the Reynolds and Maxwell stresses as
$\mathcal{Q}_{ij}=\mean{u_i u_j}$ and $\mathcal{M}_{ij}=-\mean{b_i
  b_j}/\mu_0 \mean{\rho}$, respectively. Note that
Eq.~(\ref{equ:angmom}) differs from the
formulation of, e.g., \cite{BMT04} in that we have
retained terms containing the mass flux $\rho\bm{U}$ in the
contributions corresponding to the meridional circulation. This is
because in our fully compressible setup, $\DIV{\rho\bm{U}}$ is
non-zero. This is particularly important for the average
radial mass flux $\mean{\rho U_r}$. However, we found that the effects
of compressibility in the Reynolds stress and the viscous terms are
negligible\footnote{This means that terms of the form $\mean{(\rho
    u_i)'u_j}$, where the prime denotes fluctuations, $\mean{\rho u_i
    u_j}$, and $\nu\mean{\rho u_{i;j}}$ are neglected.}.

\begin{figure*}
\begin{center}
\includegraphics[width=\textwidth]{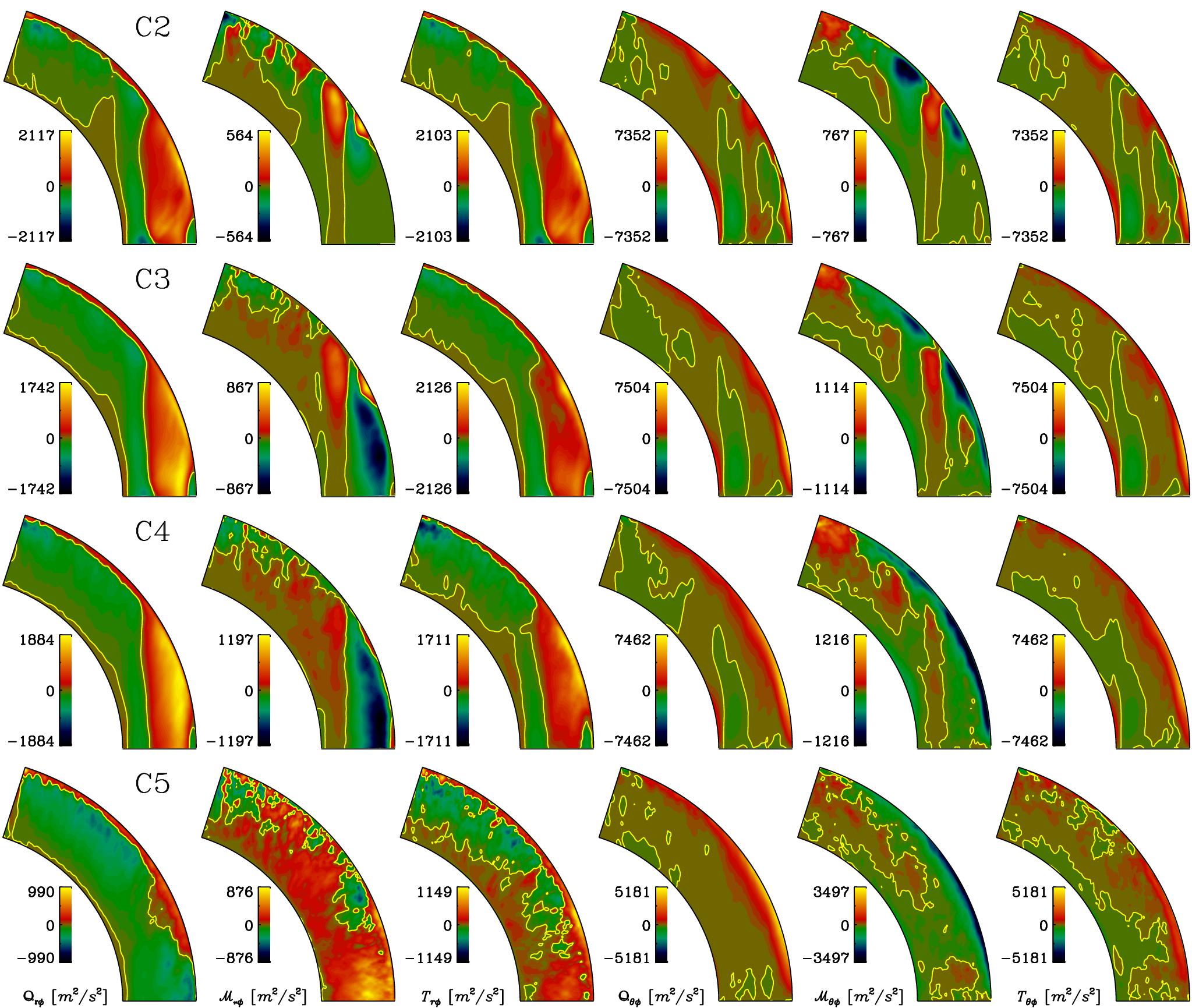}
\caption{{\it From left to right:} Reynolds, Maxwell, and total
    stress components $\qqrp$, $\mathcal{M}_{r\phi}$, $T_{r\phi}$,
    $\qqtp$, $\mathcal{M}_{\theta\phi}$, and $T_{\theta\phi}$ for
  Runs~C2 ({\it top}), C3, C4, and C5 ({\it bottom}). Data
  nearer than $2.5\degr$ from the latitudinal boundaries are
  not shown, so as to emphasize the structures at lower latitudes.
  The yellow contours denote the zero levels in each panel.}
\label{fig:plot_stress}
\end{center}
\end{figure*}

The main generator of differential rotation is commonly thought to be
the Reynolds stress and in particular its non-diffusive contribution
due to the $\Lambda$ effect \citep{R80,R89}. Disentangling the
contributions
from the $\Lambda$ effect and the turbulent viscosity is currently
only possible using assumptions regarding either of the two
coefficients and computing the other from the Reynolds stress
\citep[e.g.][]{KKB14,KKKBOP15,WKKB16}. We will not attempt this here
but study the total turbulent stress
$T_{ij}=\mathcal{Q}_{ij}+\mathcal{M}_{ij}$ and the anisotropy parameters
realized in simulations with varying magnetic Reynolds number.
Figure~\ref{fig:plot_stress} shows the off-diagonal Reynolds stress
components, $\qqrp$ and $\qqtp$, from Runs~C2 to C5 with $\ReM$ varying
between 14 and 146. The spatial structures of both stress components
remain relatively similar in Runs~C2, C3, and C4. Neither of the
stresses show a monotonous behavior as functions of $\ReM$, both being
generally smaller in Run~C3 in comparison to Runs~C2 and C4. However,
in Run~C5 the overall magnitude is decreased by a factor of two for
$\qqrp$ and about 15 per cent for $\qqtp$. No
corresponding decrease is observed in the diagonal components of the
stress, whose overall magnitude is given by the fluid Reynolds number;
see \Table{tab:runs}.

We find that for magnetic Reynolds numbers up to roughly 30, the
off-diagonal Maxwell stresses are smaller than the
corresponding Reynolds stresses. The magnitude of the vertical
component $\mathcal{M}_{r\phi}$ is between a quarter and a half of
$\mathcal{R}_{r\phi}$, whereas for the horizontal component the
difference is greater. With higher $\ReM$, the Maxwell stresses attain
similar profiles as the Reynolds stresses but with opposite signs. The
magnitudes of the Maxwell stresses increase with $\ReM$ such that they
become comparable to and locally even larger than the Reynolds
stresses at the highest magnetic Reynolds
numbers. The total vertical stress $T_{r\phi}$ decreases for
Runs~C2--C4 such that the effect is most clear near the equator. For
Run~C5 the vertical stress is dominated by the Maxwell stress near the
equator. The horizontal component $T_{\theta\phi}$ decreases
monotonically as $\ReM$ increases. The near cancellation of the total
stress at high $\ReM$
is likely to contribute significantly to the quenching of
differential rotation. These results are apparently at odds with
those of \cite{KKKBOP15} who reported that the Maxwell stress is
an order of magnitude smaller than the Reynolds stress. However,
their simulations were in a regime near the transition between
anti-solar and solar-like rotation profiles with relatively weak
and intermittent large-scale dynamos.

\begin{figure}
\begin{center}
\includegraphics[width=0.7\columnwidth]{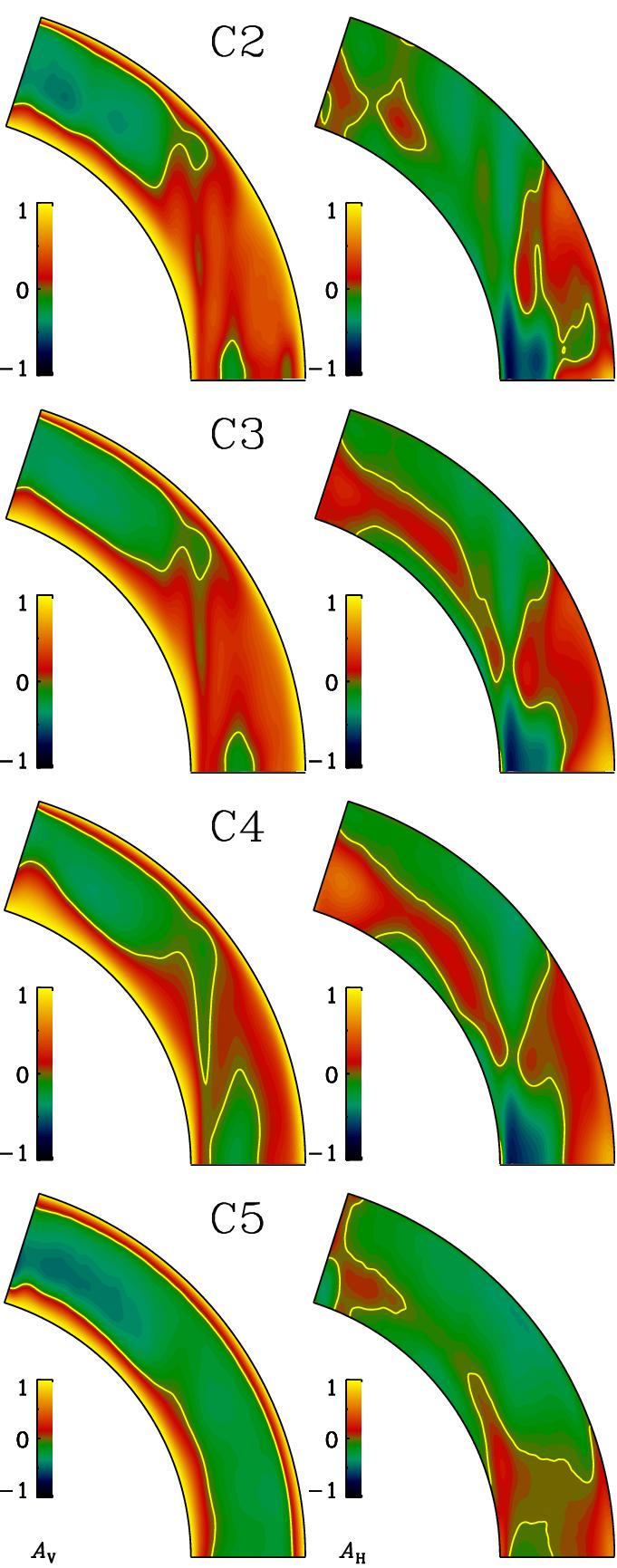}
\caption{Anisotropy parameters $A_{\rm V}$ ({\it left}) and
    $A_{\rm H}$ ({\it right}) for the same runs as in
    \Fig{fig:plot_stress}. The yellow contours denote the zero levels
    in each panel.}
\label{fig:plot_aniso}
\end{center}
\end{figure}

In Set~F with anti-solar differential rotation the Reynolds stresses
show only a weak decreasing trend as a function of $\ReM$ with a
corresponding increase in the Maxwell stress (not shown). The vertical Maxwell
stress has an opposite sign in comparison to the Reynolds stress in
all cases while a similar tendency for the horizontal stress is not so
clear. The magnitude of the vertical Maxwell stress is roughly half of
the corresponding Reynolds stress while the amplitude of the
horizontal Maxwell stress is significantly weaker than the horizontal
Reynolds stress. The much weaker quenching of the total turbulent
stress in Set~F is consistent with a clearly milder decrease of the
differential rotation than in the other sets.

Assuming that the turbulent viscosity follows a
naive mixing length estimate, the quenching of the differential
rotation can indicate that the
$\Lambda$ effect is more severely quenched by the large-scale magnetic
field at high magnetic Reynolds numbers. Another possibility is 
that small-scale magnetic fields generated
by an efficient small-scale dynamo contribute to enhancing the turbulent
viscosity. In a recent paper, \cite{HRY16} suggest that the
small-scale dynamo at high $\ReM$ suppresses velocity at small scales,
and facilitates
the growth of the large-scale magnetic fields.
First-order smoothing estimates for isotropic and homogeneous
turbulence (M.\ Rheinhardt,
private communication) suggest that turbulent viscosity acquires a
contribution $\nut^{\rm (SSD)} = \onethird b_{\rm rms}^{(0)}
\kf^{-1}$, where $b_{\rm rms}^{(0)}$ is the rms-value of the
saturated fluctuating magnetic field due to small-scale dynamo.
We find that
the fluctuating magnetic field energy grows monotonically as a
function of $\ReM$; see \Table{tab:energies}. A corresponding increase
of turbulent viscosity would be compatible with a strong decrease of
differential rotation. 
However, we find two counter-examples: in Set~A,
no SSD is present but the differential rotation still experiences
strong quenching, and in Set~F, an SSD is present but differential
rotation remains strong.
Furthermore, the dependence of the $\Lambda$ effect on small-scale
magnetic fields is currently unknown.
Thus the question of the effect of
small-scale dynamo on the turbulent transport of angular
momentum at large scales remains
open.

We also show the anisotropy parameters (\Fig{fig:plot_aniso})
\begin{eqnarray}
A_{\rm V} &=& \frac{\qqpp-\qqrr}{\qqpp+\qqrr},\\
A_{\rm H} &=& \frac{\qqpp-\qqtt}{\qqpp+\qqtt},
\end{eqnarray}
which are proportional respectively to the vertical
and horizontal $\Lambda$ effects in mean-field
hydrodynamics \citep{R80} under the assumption of slow rotation
\citep[see also the numerical results of][]{KB08}. Again the
differences between Runs~C2 and C3 are relatively small, whereas in
Runs~C4 and C5 the mostly positive $A_{\rm V}$ at mid-latitudes in
lower-$\ReM$ runs gives way to negative values. According to
mean-field theory, this corresponds to a sign change of the vertical
$\Lambda$ effect that is responsible for generating radial
differential rotation \citep{R80}. The magnitude and the spatial distribution of
the horizontal anisotropy parameter $A_{\rm H}$ are significantly
different in Run~C5 in comparison to the other runs whereas the
differences between the other runs are less significant. The negative
values of
$A_{\rm H}$ at mid-latitudes in Runs~C2--C4 coincide with the minimum
in the angular velocity suggesting that the horizontal $\Lambda$
effect is negative there. In Run~C5 the overall
magnitude of $A_{\rm H}$ is diminished near the surface at low
latitudes. Along with the change of sign of $A_{\rm V}$ this is likely
to contribute to the reduced differential rotation as a function of
$\ReM$. However, more theoretical work is needed to disentangle the
effects of the small- and large-scale magnetic fields on the angular
momentum transport.

\begin{figure}
\includegraphics[width=\columnwidth]{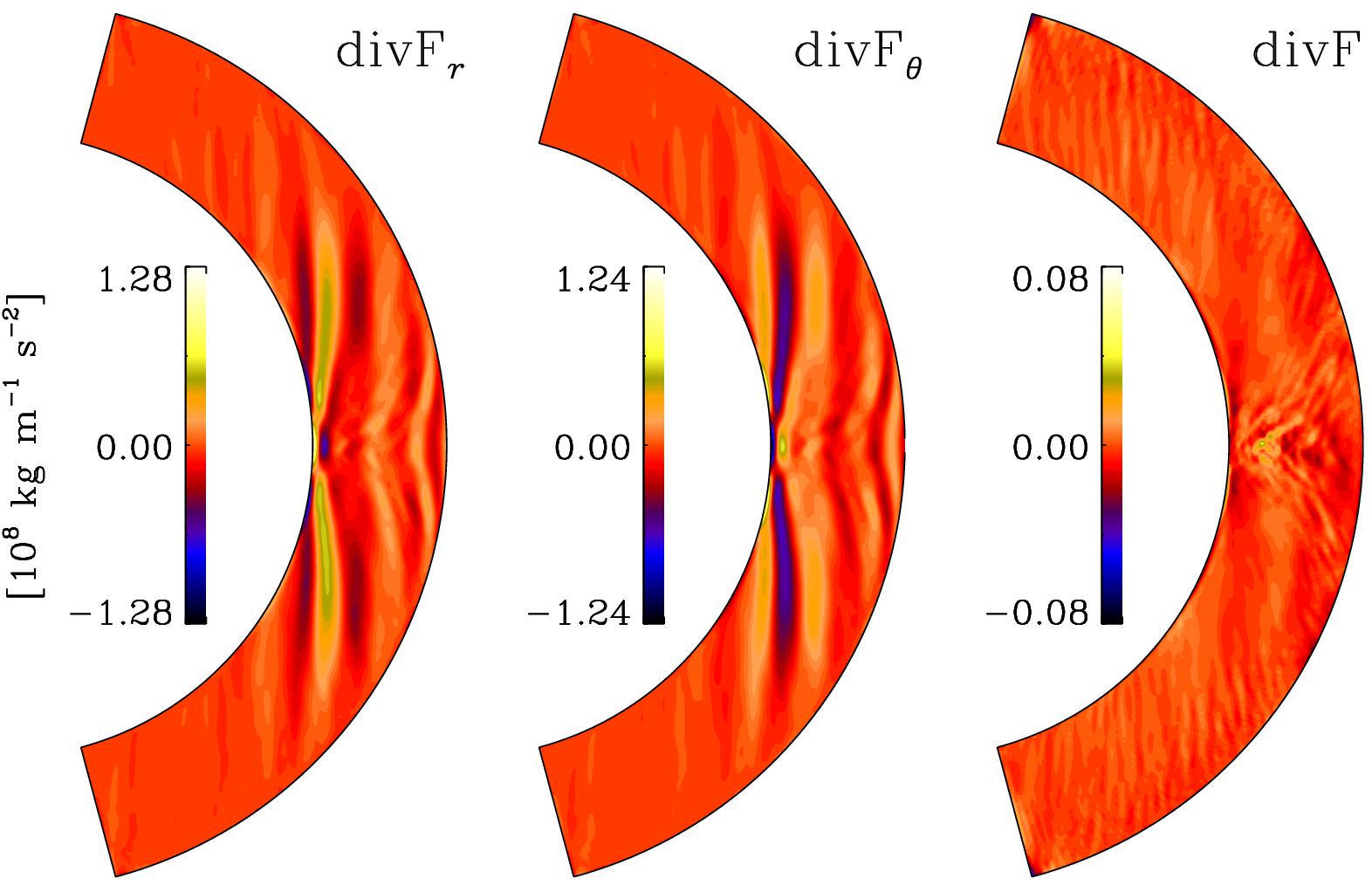}
\caption{Radial ($\mbox{divF}_r$, \emph{left panel}) and latitudinal
  ($\mbox{divF}_\theta$, \emph{middle}) parts of the divergence of the
  angular momentum fluxes. The \emph{right panel} shows the total
  divergence. The units are given in the legend. Data taken from
  Run~C3.}
\label{fig:pangflux_C3}
\end{figure}

The discussion above is valid, if the angular momentum
is in a statistically steady state demanding
that the sum of two terms, $\mbox{divF}_r + \mbox{divF}_\theta$, 
vanishes, i.e.,
\begin{eqnarray}
  \frac{1}{r^2} \frac{\pd (r^2\mathcal{F}_r)}{\pd r} + \frac{1}{r\sin\theta}\frac{\pd (\sin\theta \mathcal{F}_\theta)}{\pd \theta}\equiv \mbox{divF}_r + \mbox{divF}_\theta = 0,
\label{equ:divF}
\end{eqnarray}
where
\begin{equation}
\mathcal{F}_r\!=\!\varpi\bigg[\varpi \mean{\rho U_r}\Omega\!+\!\mean\rho\mathcal{Q}_{r\phi}\!-\!\nu \rho \varpi \frac{\pd \Omega}{\pd r}\!-\!\mu_0^{-1}(\mean{B}_\phi\mean{B}_r\!+\!\mean{b_r b_\phi})\bigg],
\end{equation}
and
\begin{equation}
\mathcal{F}_\theta\!=\!\varpi\bigg[\varpi \mean{\rho U_\theta}\Omega\!+\!\mean\rho\mathcal{Q}_{\theta\phi}\!-\!\nu \rho \frac{\varpi}{r} \frac{\pd \Omega}{\pd \theta}\!-\!\mu_0^{-1}(\mean{B}_\phi\mean{B}_\theta\!+\!\mean{b_\theta b_\phi})\bigg].
\end{equation}
Figure~\ref{fig:pangflux_C3} shows representative results
from Run~C3 for the terms $\mbox{divF}_r$ and $\mbox{divF}_\theta$,
as well as their sum.
We find that the radial and latitudinal contributions have
similar structures but opposite signs and that their sum is at most
roughly seven per cent of the individual components. The elongated
structures at low latitudes within the tangent cylinder are due to the
meridional circulation which yields the dominant contribution to the
divergence. We note that neither the radial nor the latitudinal
fluxes need to individually cancel for the divergence to vanish.

\subsection{Large-scale magnetic fields and dynamo cycles}

\subsubsection{General considerations}
We find no dynamos in the lowest magnetic Reynolds number cases
Runs~A1, A2, B1, C1, and D1 which are in the range $\ReM=5\ldots9$. On
the other hand, the highest magnetic Reynolds numbers clearly exceed
critical values for small-scale dynamo action to occur in simpler
setups \citep[e.g.][]{Schekoea05}. In the present simulations the
small- and large-scale dynamos
can be excited at the same time and disentangling the two is not
directly possible. Thus we resort to runs where we artificially
suppress the large-scale ($\phi$-averaged) magnetic fields at each
time step. This eliminates the large-scale dynamo and growing magnetic
fields can be associated with a small-scale dynamo. We have performed
such runs for each of the cases where a dynamo is observed and find
that a small-scale dynamo is excited in Runs~B5, C5, D4, D5, E3, E4,
F4, G2, and G3.

\begin{table}[t!]
\centering
\caption[]{Cycle lengths detected using $D^2$ statistic.}
       \label{tab:d2}
      $$
          \begin{array}{p{0.05\linewidth}ccccccc}
            \hline
            \hline
            \noalign{\smallskip}
\multirow{2}{*}{Run} & \multirow{2}{*}{$P^{\rm max}$} &
\multirow{2}{*}{$l_{\rm coh}^{\rm max}$} &
\multicolumn{4}{c}{\mbox{Cycle length}} \\ & & & \mbox{$\mean{B}_{\phi}$(N)}
& \mbox{$\mean{B}_{\phi}$(S)} & \mbox{$\mean{B}_r$(N)} & \mbox{$\mean{B}_r$(S)} &
\mbox{Class}\\
\hline
A3 & 11        & 5         & 4.06      & -         & 3.71      & -     & \mbox{PW}   \\ 
A4 & 7         & 5         & -         & 5.31      & -      & -     & \mbox{PW}   \\ 
A5 & 5         & 5        & -         & -         & -         & -     & \mbox{(QS/IR)}    \\ 
\hline
B2 & 5         & 5        & 1.57      & 1.42      & 2.29      & 1.92  & \mbox{PW}    \\ 
B3 & 5         & 5         & -         & -         & 4.57      & 4.77  & \mbox{IR}    \\ 
B4 & 10        & 5         & 8.14      & -         & 9.16      & -     & \mbox{IR}    \\ 
B5 & 5         & 5        & -         & -         & 4.90      & 4.90  & \mbox{(PW)}    \\ 
\hline
C2 & 7         & 6        & 3.40      & 3.34      & 3.40      & 3.34  & \mbox{EW}    \\ 
C3 & 21        & 5         & 5.11      & 5.24      & 4.99      & 5.11  & \mbox{EW}    \\ 
C4 & 13        & 5         & -         & 7.30      & 8.81      & 7.56  & \mbox{EW}    \\ 
C5 & 7         & 5         & -         & -         & -         & 6.80  & \mbox{(QS/IR)}    \\ 
\hline
D2 & 6         & 9         & 3.44      & 3.39      & 3.44      & 3.39  & \mbox{EW}    \\ 
D3 & 12        & 5         & 5.85      & 5.27      & 5.70      & 5.14  & \mbox{EW}    \\ 
D4 & 8         & 5        & -         & -         & -         & -     & \mbox{(EW)}    \\ 
D5 & 5         & 5         & -         & -         & -         & -     & \mbox{(QS/IR)}    \\ 
\hline
E1 & 2         & 5         & 1.66      & 1.50      & 1.70      & -     & \mbox{PW}    \\ 
E2 & 11        & 5         & 5.89      & 6.05      & 6.23      & 6.05  & \mbox{IR}    \\ 
E3 & 14        & 5         & 8.46      & 6.42      & 8.81      & 6.05  & \mbox{(QS/IR)}    \\ 
E4 & 4         & 5         & -         & -         & -         & -  & \mbox{(QS/IR)}    \\ 
\hline
G2 & 10        & 5        & 5.96      & -         & 6.30 (9.57) & 9.57 (6.48) & \mbox{EW/QS} \\ 
\hline
          \end{array}
          $$
\tablefoot{Here $P^{\rm max}$ and $l_{\rm coh}^{\rm max}$ are the
  upper limit for the period search range and the maximum coherence
  length, respectively. They are related via $\Delta t \approx P^{\rm
    max} l_{\rm coh}^{\rm max}$. Values in the fourth to seventh
  columns are detected cycle periods (in years) for azimuthal
  ($\mean{B}_\phi$) and radial ($\mean{B}_\phi$) mean magnetic fields
  in
  north (N) and south (S) hemispheres. ``--" stands for no detection.
  The last column indicates the classification of the dynamo solution,
  see Sect.~\ref{sec:dynmod}. Brackets indicate an uncertain
  classification due to too short time series.}
\end{table}

\subsubsection{Cycle detection using $D^2$ analysis}\label{D2res}

Using the $D^2$ statistic discussed in \Sec{DataAnalysis},
we analyzed separately radial and azimuthal components of
the mean magnetic field. In the former case we included all the data
near the polar region ($55^{\circ} \le |90\degr-\theta| \le
75^{\circ}$) and in the latter case data around mid-latitude region
($10^{\circ} \le |90\degr-\theta|
\le 45^{\circ}$). We considered data above $0.94R_\odot$ and analyzed
the northern and southern hemispheres
separately.

\begin{figure*}
\begin{center}
  \includegraphics*[width=0.5\textwidth]{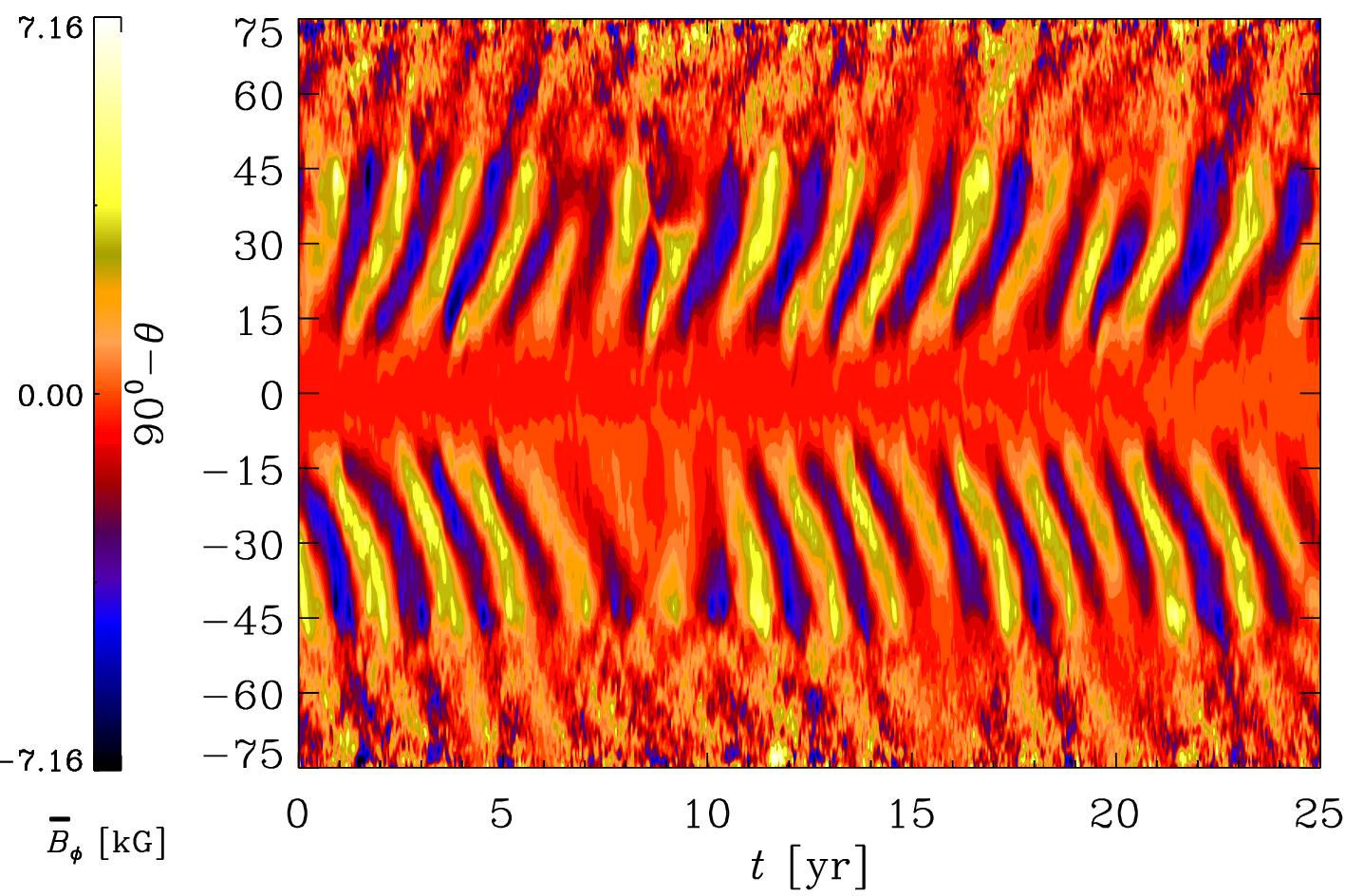}\includegraphics*[width=0.5\textwidth]{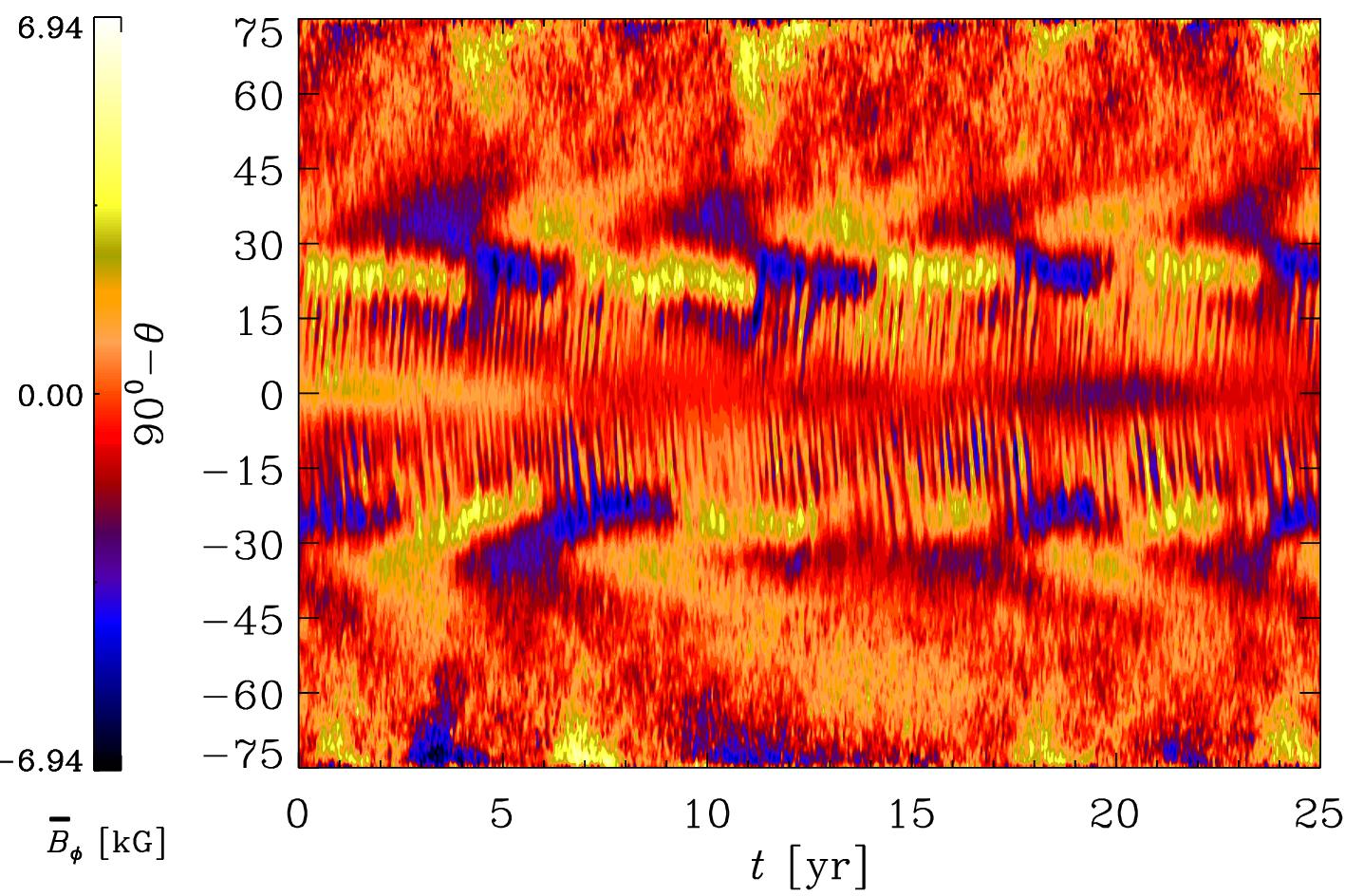}
  \includegraphics*[width=0.5\textwidth]{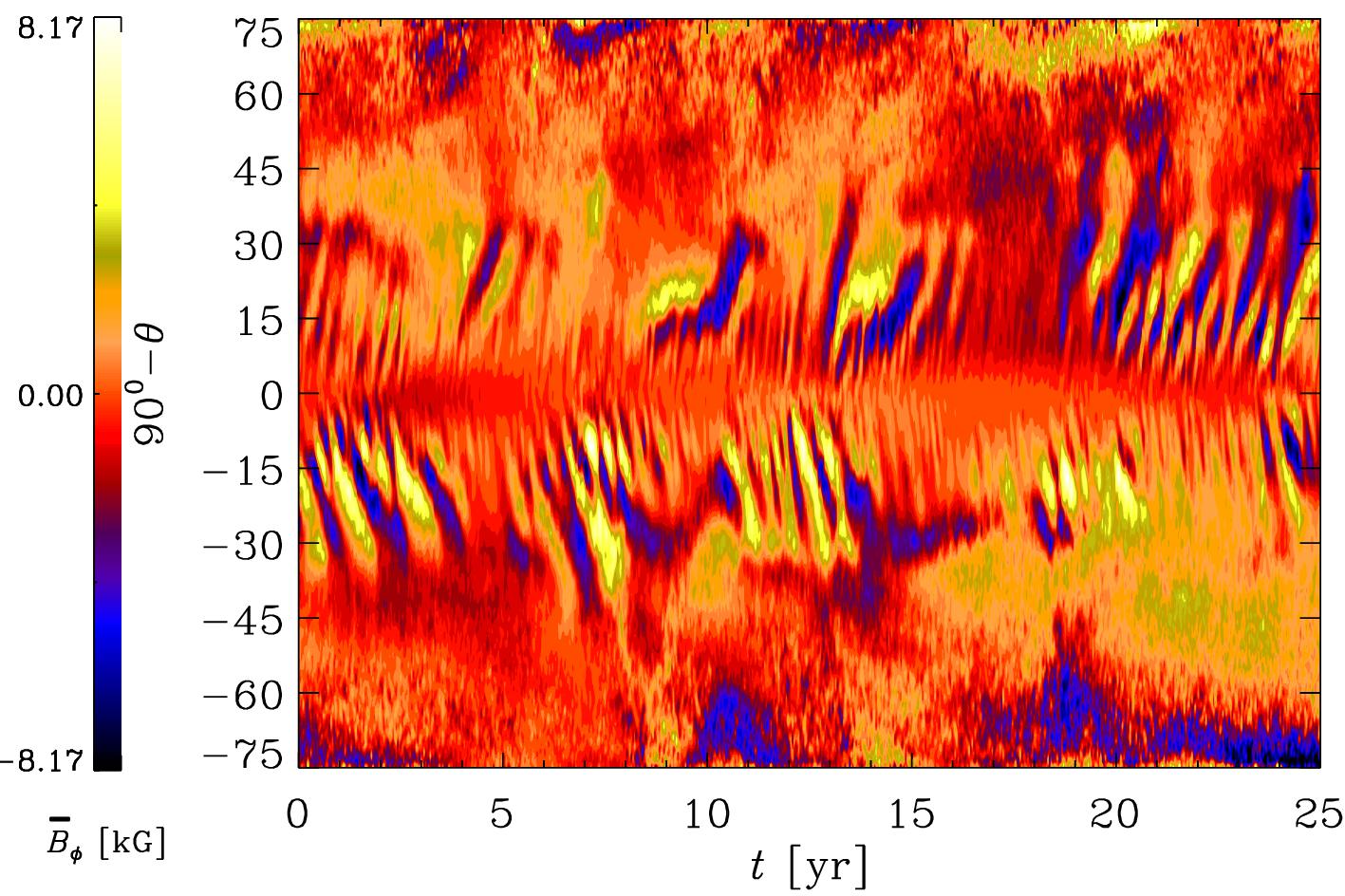}\includegraphics*[width=0.5\textwidth]{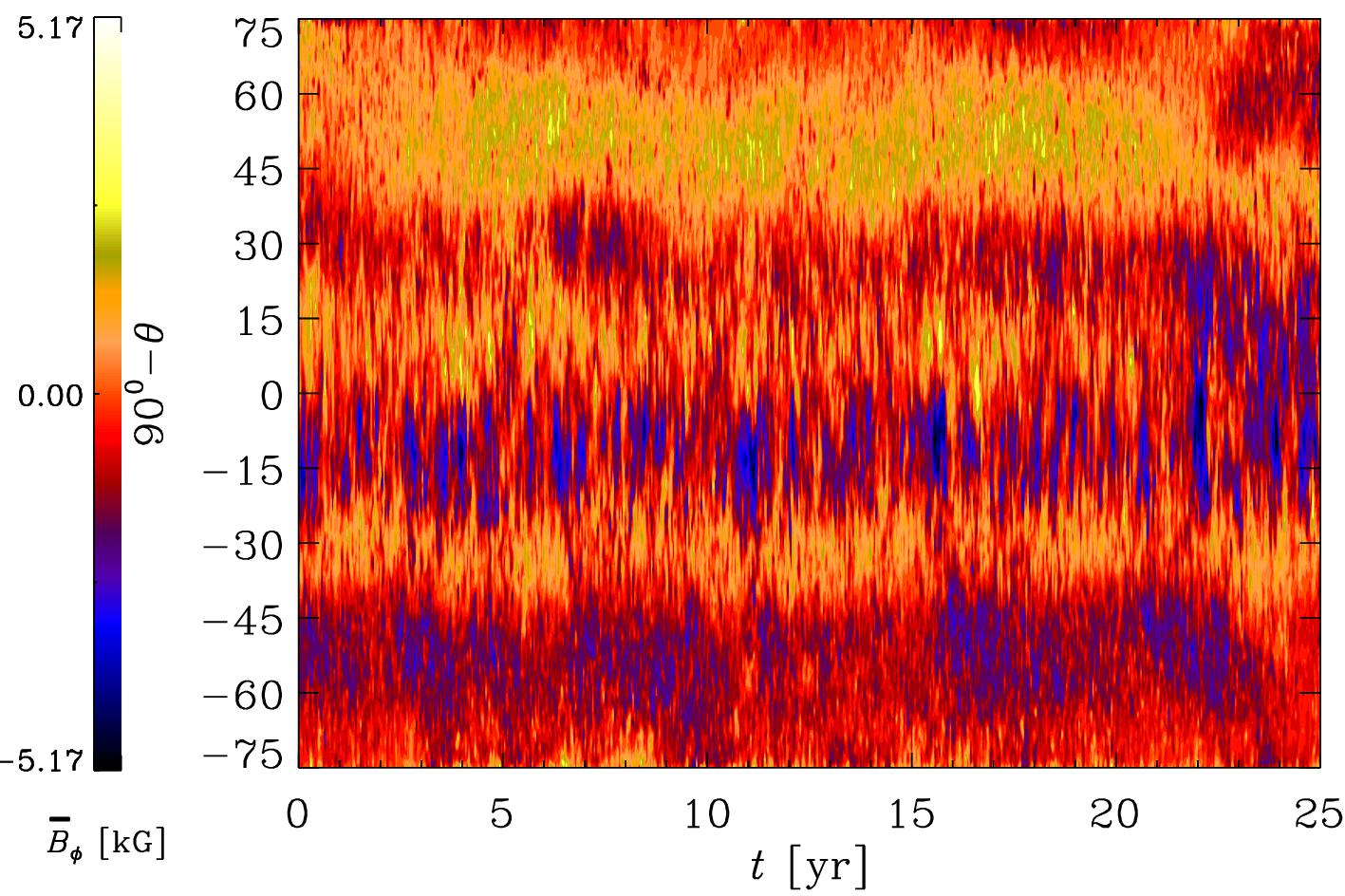}
\end{center}
\caption{Azimuthally averaged azimuthal magnetic field near the
  surface from Runs~B2 ({\em top left}), D3 ({\em top right}), B3
  ({\em bottom left}), and F3 ({\em bottom right}) with poleward (PW),
  equatorward (EW), irregular (IR), and quasi-stationary (QS)
  dynamo solutions, respectively.}
\label{fig:pbutter}
\end{figure*}

The results of the analysis are listed in \Table{tab:d2}.
We required that at least five full cycles were covered for each trial period
in the period search range. To meet this criteria we adjusted the upper limit of the period search range
according to the length of the dataset, while the lower limit was always fixed at one year.
We set the lower limit for the coherence length range to two cycles
per given period and the upper value was determined by the dataset
length.
However, in some cases the longest possible trial period was still
below two years in which cases the analysis was deemed infeasible.
This situation was encountered for Run G3, which has therefore not been included in
\Table{tab:d2}. Furthermore, Runs~A1, A2, B1, C1, and D1 without
dynamos were not analysed.

All the cycle lengths given in the table are
significant with the p-values less than 1\%. More precisely, in our case the p-value represents a probability that a cycle
with a given period would appear by chance out of white noise data
with the same distribution as the original data.
In those cases where the obtained $D^2$ spectrum
contained no significant minima, no cycle was detected.
However, here we must also note that due to the narrowing of the period
search range in some cases it is possible that the real cycle length
is located outside the range and we did not detect it.
This is particularly relevant for the runs at high magnetic Reynolds
numbers where the time series are short.

In all the sets, especially if one leaves out the highest Reynolds number cases, the cycle
lengths are found to increase as functions of the magnetic Prandtl number. Thus, the
dynamo period is sensitive to the strength of the magnetic diffusion such
that when the diffusion is decreased, and correspondingly the diffusion time scale increases, the
dynamo period gets longer.  

\subsubsection{Dynamo modes}
\label{sec:dynmod}

Given that recent simulations reproduce solar-like magnetic cycles with
equatorward migration (hereafter EW) of activity belts, it is of
interest to probe the parameter space to determine when such solutions
are excited.
Having identified cyclic solutions with the $D^2$ statistics we
classify them as EW or PW (poleward) based on the migration direction
of the dynamo wave at low latitudes between $\pm 10\degr$ and
$\pm45\degr$. We also find quasi-stationary (hereafter QS) solutions
in Set~F. If a clear classification by visual inspection cannot be
made we classify the solution as irregular (IR). In some cases
features from more than one class can be present, e.g.\ an equatorward
cyclic variation on top of a quasi-stationary background. This
solution is classified as EW/QS but typically such cases are quite
uncertain and are indicated by brackets in the last column of
\Table{tab:d2}.
No large-scale dynamo action is denoted by ND.
In Fig.~\ref{fig:pbutter} we show time-latitude plots of the
azimuthally averaged azimuthal magnetic field $\mean{B}_\phi$ from
four runs exemplifying each of the main dynamo modes discussed above.

We first consider the Sets~A--D with $\Omega_0=5\Omega_\odot$.
For $\PraSGS=0.25$ (Set~A) and moderate $\ReM$ we obtain
solutions with clear poleward migration. This type of
dynamos were first obtained already in the pioneering studies of
\cite{Gi83} and \cite{Gl85}.
Solutions
showing poleward migration have been reported more recently by many
groups \citep[e.g.][]{KKBMT10,BMBBT11,SPD12}.
In one case (Run~A3) we see a hemispheric
dynamo with
poleward migration, similar to those reported by e.g.\ \cite{Bu02} and
\cite{GDW12}.
In the largest $\ReM$ case (Run~A5) the clear oscillatory solutions of
the runs with smaller Reynolds numbers give way to possibly
irregularly reversing or quasi-stationary large-scale fields. The time
series is too short, however, for possible cycles to be detected.
For $\PraSGS=0.5$ (Set~B) we find a PW mode excited in the case
$\ReM=12$ (Run~B2). At intermediate $\ReM$ (Runs~B3 and B4) the
solutions mostly show irregularly reversing fields, although features
from both PW and EW modes can be discerned at times; see the lower
left panel of Fig.~\ref{fig:pbutter}.
The highest Reynolds number case (Run~B5) appears to return to a
poleward oscillatory mode with a longer cycle period. However, the
large-scale field shows only four reversals in the 24~year duration of
the simulation and the $D^2$ statistic captures this only in one of
the four analyzed cases. Therefore
the classification of this run as PW is deemed less robust than those
of the other runs in this set.

\begin{figure}
\begin{center}
\includegraphics[width=0.45\columnwidth]{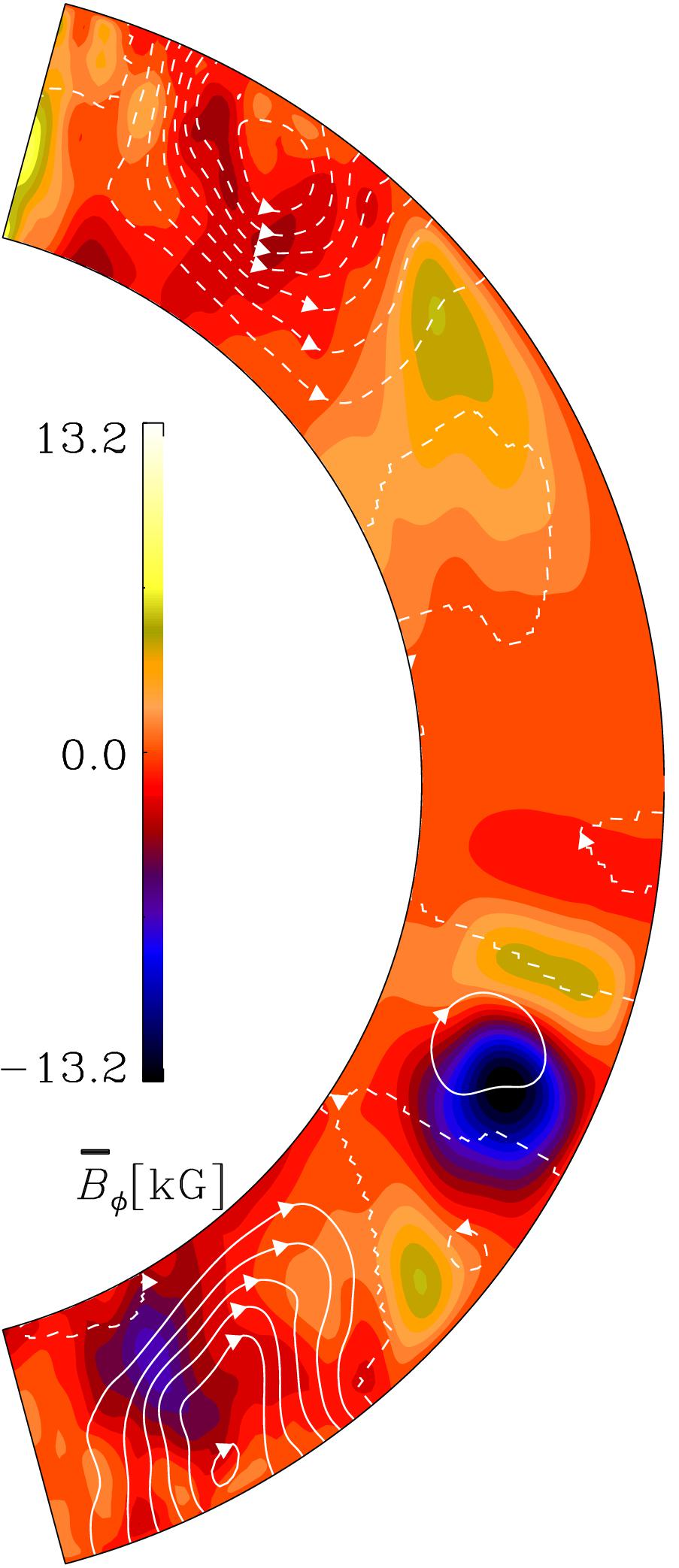}\includegraphics[width=0.45\columnwidth]{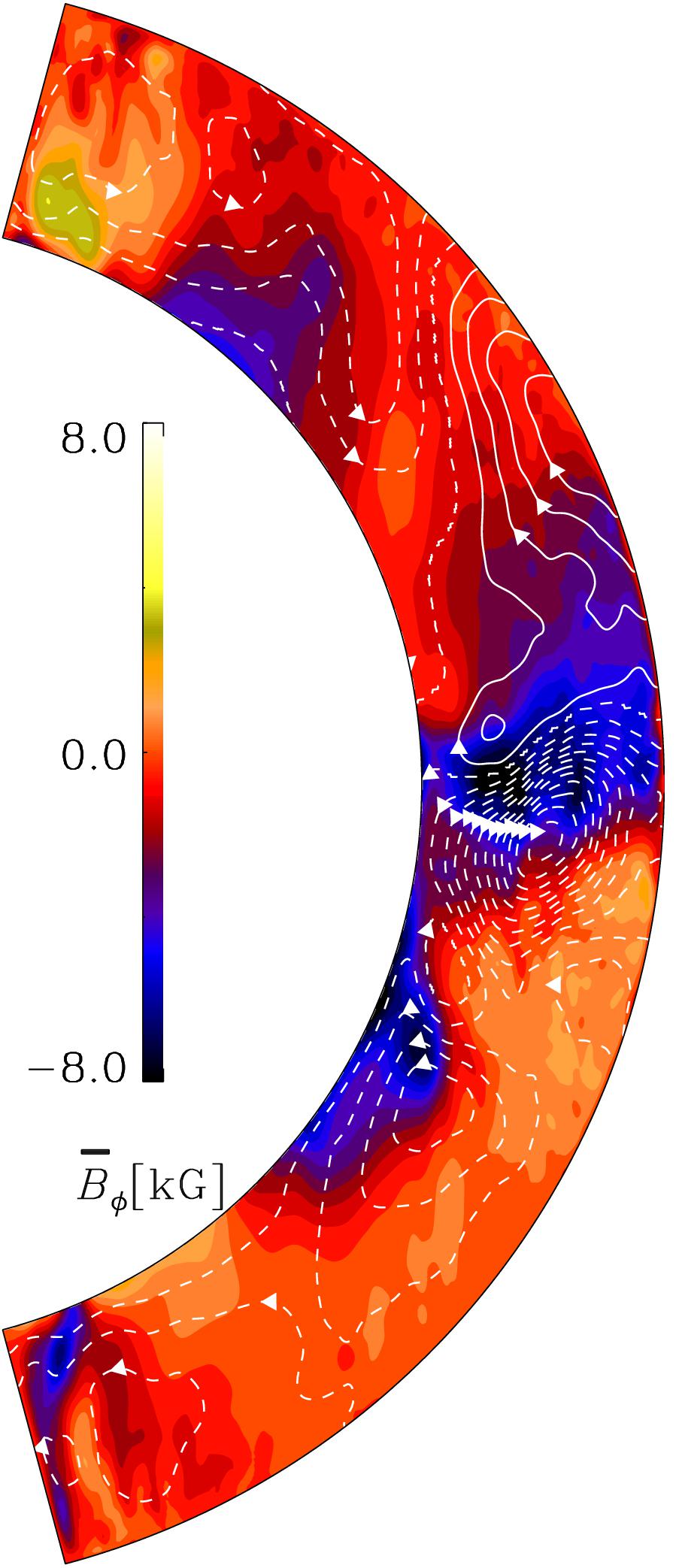}
\end{center}
\caption{Azimuthally averaged azimuthal magnetic field $\mean{B}_\phi$
  (colour contours) in units of kG and the fields lines of the
  poloidal field (continuous and broken lines for clockwise and
  anticlockwise loops, respectively). {\it Left:} Data from Run~B2
  averaged over three months near a cycle maximum at
  $t=70$~years. {\it Right:} Data from Run~A5 averaged over the last
  10 years of the run.}
\label{fig:pfields_A5}
\end{figure}

\begin{figure*}
\sidecaption
\includegraphics[width=12cm]{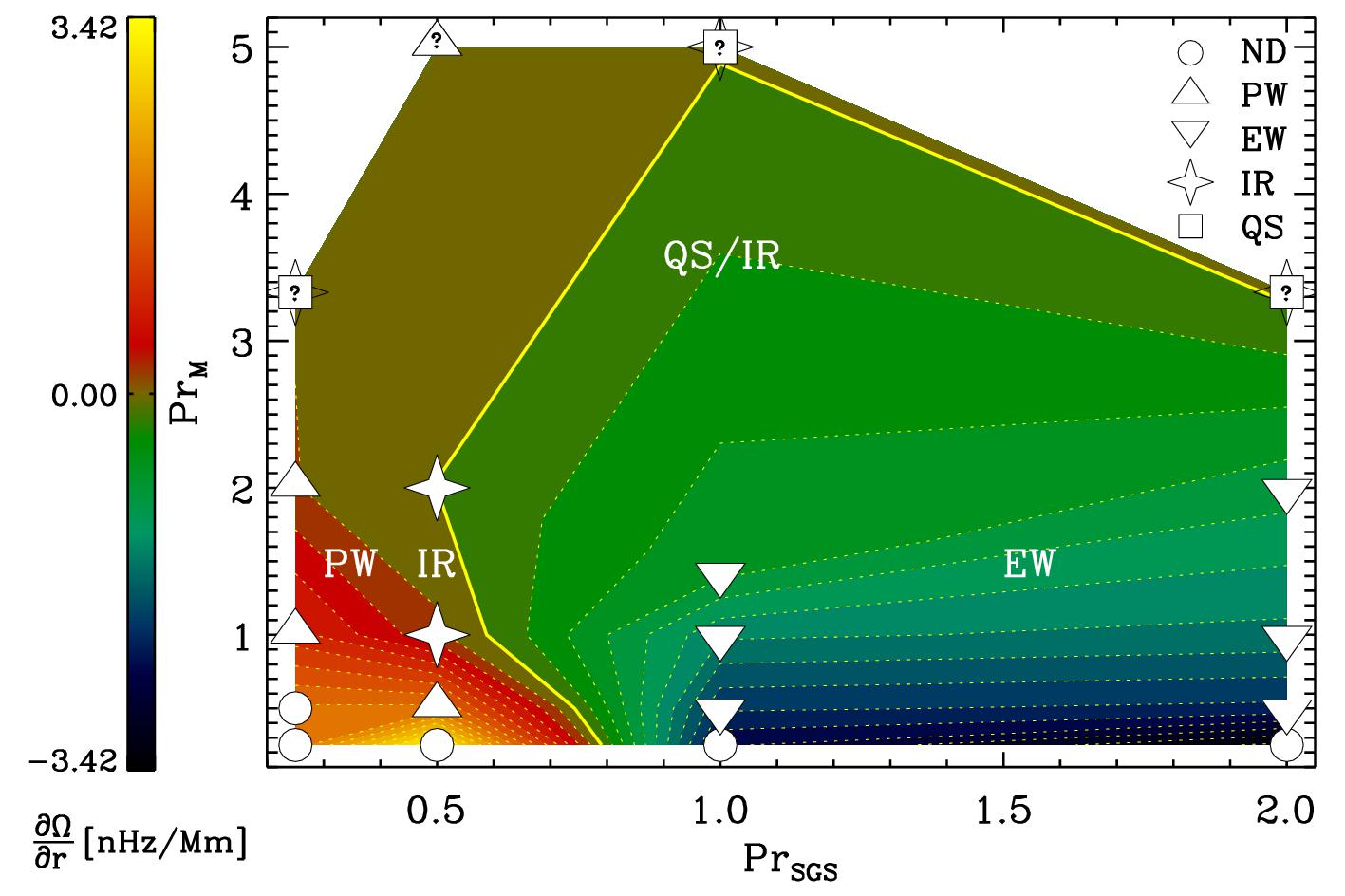}
\caption{Colour contours: radial derivative of $\Omega$ at
  $r=0.85R_\odot$ averaged from latitudes $+25\degr$ and $-25\degr$.
  The dynamo modes realized in Sets~A--D are overplotted with
  abbreviations ND, PW, EW, IR, and QS. The thick yellow curve
  indicates the zero level of $\pd\Omega/\pd r$. Overlapping symbols denote
  solutions that show characteristics from several modes while question marks
  indicate that the classification is uncertain due to insufficient length of
  the data set.}
\label{fig:pdmodes}
\end{figure*}

For $\PraSGS\ge1$ (Sets~C and D) we often find solutions with EW
migration (Runs~C2, C3, D2, D3, and D4). In Run~D4 the solution is
clearly EW but the time series covers only three full cycles and is
thus not detected by $D^2$. The transition from poleward
migrating to equatorward
migrating solutions at intermediate magnetic Reynolds numbers
coincides with the change of the rotation profile
from runs with consistently positive radial gradient of
$\Omega$ to ones with a minimum of $\Omega$ at
mid-latitudes where $\pd\Omega/\pd r < 0$ as $\PraSGS$ increases.
The change in the dynamo mode fits with the interpretation in terms of
a classical dynamo wave obeying the Parker--Yoshimura rule
\citep{WKKB14,WRKKB16}.
For the highest magnetic Reynolds number cases in Sets~C and D we
again observe possibly quasi-stationary or irregular configurations. This
classification is based on significantly shorter time series than in
the other cases and it is possible that there are cycles that are much
longer than in the low-$\ReM$ cases or that a prolonged transient is still
in progress.
These results appear to be in agreement with those of \cite{SPD12} who
found a transition from oscillatory dynamos to non-oscillatory ones at
a roughly comparable $\ReM$ for Boussinesq convection in spherical
shells. They also showed that in the high--$\PrM$ regime the two
bistable branches of dynamo solutions merge and that only strongly
dipolar dynamos with weak differential rotation survive. However,
\cite{GDW12} have shown that for sufficiently density stratified cases
the dipolar
branch does not exist at least for moderate values of $\ReM$. Our
simulations with moderate density stratification of $H_\rho\approx3$,
where $H_\rho=-(\pd\ln\rho/\pd r)^{-1}$ is the density scale height, do
not indicate that a dipolar mode takes over high $\ReM$, see
\Fig{fig:pfields_A5} for a representative result from Run~A5. This is
consistent with the results of \cite{GDW12} who found no
dipole-dominated dynamos for $H_\rho\gtrsim2$.

The realized solutions from the runs in Sets~A--D are plotted in the
$(\PraSGS,\PrM)$--plane in Fig.~\ref{fig:pdmodes}. 
We find that the low and intermediate $\ReM$ runs produce cyclic
solutions with PW for low $\PraSGS$ and EW for high $\PraSGS$. The
case $\PraSGS=0.5$ works as a watershed between the two cyclic
regimes. The dynamos at high $\ReM$ are fundamentally different from
their more diffusive counterparts in that the differential rotation is
almost absent. Although the ratio of the energies in the mean poloidal
to toroidal components is not significantly different in the
high-$\ReM$ runs in comparison to lower-$\ReM$ runs in each set, the
dynamos at the high--$\ReM$ regime can be of $\alpha^2$ type.
Substantiating this claim, however, requires that the turbulent
transport coefficients relevant for the maintenance of large-scale
magnetic fields are extracted from the simulations and applied in
corresponding mean-field models which is not within the scope of the
present study.

\begin{figure*}
\begin{center}
\includegraphics[width=0.8\textwidth]{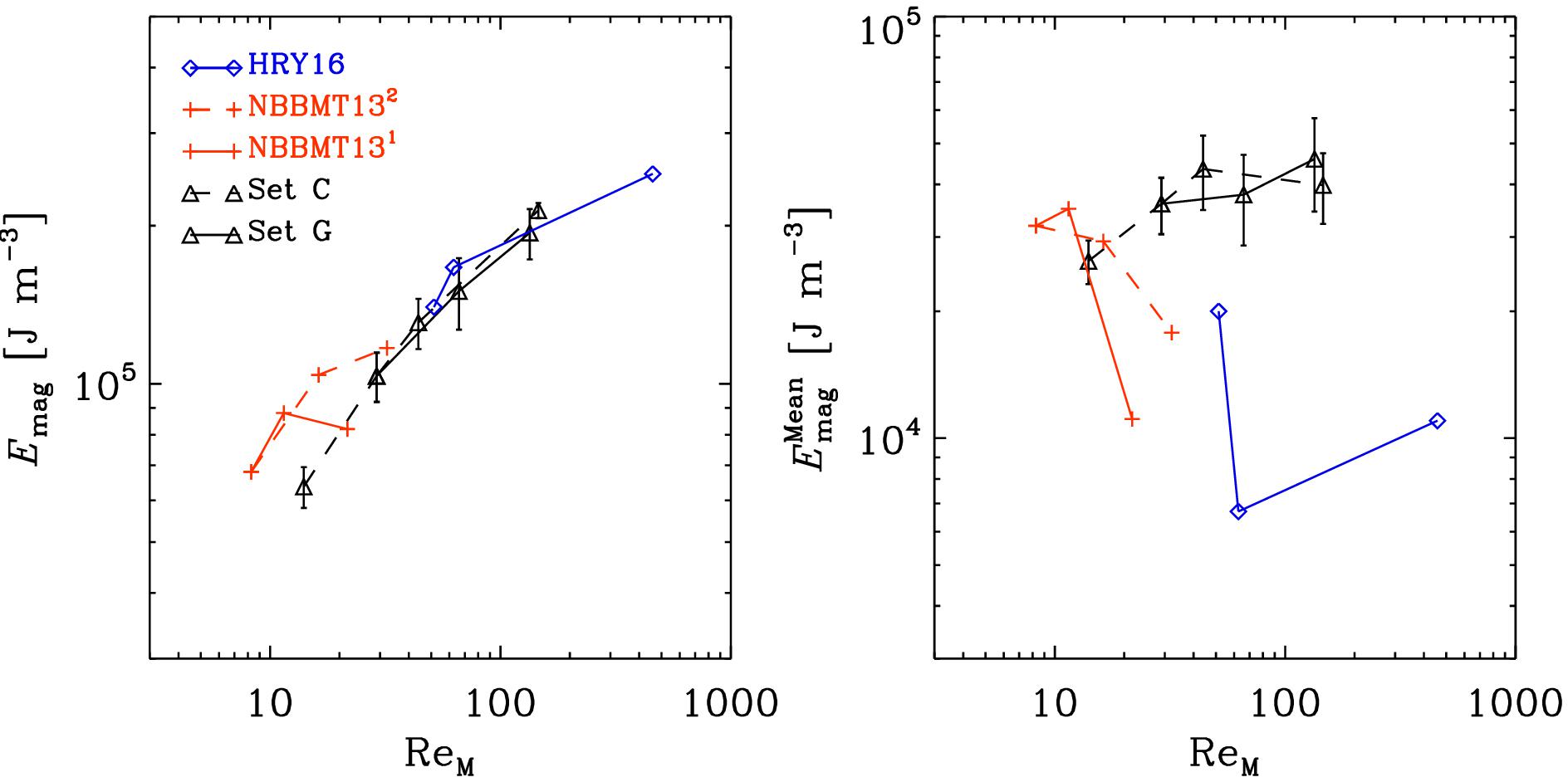}
\end{center}
\vspace{-0.5cm}
\caption{Energy densities of the total magnetic field ({\em left
    panel}), and the azimuthally averaged fields ({\em right}) from
  Sets~C (black dashed line) and G (black solid), from two sets of
  runs by \cite{NBBMT13} (solid and dashed red lines) and from
  \cite{HRY16} (blue solid lines). The red solid line consists of data
  from Cases~D3, D3a, and D3b of \cite{NBBMT13} with
  $\PrM=\mbox{const}=0.5$. Correspondingly, the dashed red line shows
  data from Cases~D3, D3-pm1, and D3-pm2 of \cite{NBBMT13} with $\PrM$
  varying from $0.5$ to $2$. From \cite{HRY16} we show data for Cases
  `Low', `Medium', and `High'.}
\label{fig:pene_comp}
\end{figure*}

In Set~E we find a PW solution for the lowest $\ReM$ (Run~E1)
whereas at higher $\ReM$ we find IR (E2) and irregular/quasi-stationary
(E3 and E4) configurations. Generally, Set~E behaves similarly as
Sets~A--C,
i.e.\ a transition from oscillatory dynamos at intermediate $\ReM$ to
quasi-stationary or irregularly varying solutions at high $\ReM$.
However, Run~B1 of \cite{WKKB16} is similar to Run~E2, but with
$\PraSGS=2$ instead of $1$ shows EW.
Set~F
is qualitatively different from the other sets similarly as for the
differential
rotation. The large-scale magnetic fields show a quasi-stationary
configuration in all of the runs in Set~F.
In Run~G2 the modulation of the differential rotation was shown to be
related to a changing large-scale dynamo mode that is either EW or QS;
see panels (b) and (d) of \Fig{fig:penet}. In the highest-$\ReM$ case,
Run~G3, the two competing modes appear to be present again, but the
short data set length renders such classifications preliminary at
best.

\subsubsection{Saturation level of large-scale magnetic fields}
\label{sec:satu}

For the total magnetic energy we find a monotonically increasing
trend as a function of $\ReM$ in all sets, see representative results
in the left panel of \Fig{fig:pene_comp} and
Table~\ref{tab:energies}. The absolute value of the
axisymmetric parts of the poloidal and toroidal magnetic fields shows
monotonically decreasing trends only in Set~F with anti-solar
differential
rotation. In the other sets
the energies of the poloidal and toroidal mean fields typically do not
behave monotonically as functions of $\ReM$. However, given the large
temporal variations, our results are compatible with mean magnetic field
energies converging to constant values at high magnetic Reynolds
numbers, see the right panel of \Fig{fig:pene_comp} for the results of
Sets~C and G. The same conclusion applies also to Set~F. The findings
for the mean magnetic fields appear to be in
contradiction with the results of \cite{NBBMT13} who found a
monotonically decreasing trend as a function of $\ReM$; see
\Fig{fig:pene_comp}.

The main difference to the simulations of \cite{NBBMT13}
is that their models were done with a full spherical shell as
opposed to the wedge geometry used here, and that magnetic
field boundary condition at the outer radial boundary uses
extrapolation to a potential field rather than a radial field
condition as in the
present study.
Furthermore, \cite{NBBMT13} consider anelastic models whereas in our
case the gas is fully compressible.
Simulations with forced turbulence in spherical shells with coronal
envelopes have shown that a higher field strength can be achieved, if
the magnetic field at the boundary is not restricted to being purely
radial \citep{WB14}.
Furthermore, \cite{NBBMT13} use profiles for $\nu$ and
$\eta$ which are absent in our study. The constant diffusion
coefficients used here may lead to a steeper increase of the magnetic
field energy than in the runs of \cite{NBBMT13} where the local value
of $\ReM$ is significantly higher in the deeper layers.
Detailed comparisons of diffusion schemes and parameter values
  are presented in Tables~\ref{tab:methods} and \ref{tab:values} in
  Appendix~\ref{sec:comp}.

It is difficult to assess which of these differences is the most
important. However, a plausible candidate is the change of topology of
the field in the full-sphere simulations of \cite{NBBMT13}. As
described in \cite{KMCWB13},
the large-scale magnetic field can become dominated by low-order
non-axisymmetric modes and therefore applying an axisymmetric mean
will average out such field contributions (see their Fig.~16 and Table
2). Evidence of such non-axisymmetric modes can be seen in the
instantaneous magnetic fields in Figs.~4 and 6 of
\cite{NBBMT13}. However, it is not possible to assess how the degree
of non-axisymmetry behaves as a function of $\ReM$ in the results of
\cite{NBBMT13} and thus a direct comparison is not possible.

In another recent study, \cite{HRY16} present results from less
rapidly rotating ($\Omega=\Omega_\odot$) simulations at high values of
$\ReM$. Such simulations are less likely to contain significant
non-axisymmetric modes. Their main claim is that whereas the mean
magnetic energy decreases at intermediate $\ReM$, it recovers as $\ReM$
is increased further. We have included their results in
\Fig{fig:pene_comp} for comparison. However, only a rough comparison is
possible as only
three data points are available and a jump occurs between the two
lowest-$\ReM$ runs which is caused by a switch from explicit diffusion
to a numerical slope-limited scheme. The absolute values for the
  mean fields are also
lower than ours likely because their large-scale dynamo is less
efficient than in the present study due to their slower rotation. Thus
only the trend as a function of $\ReM$ can be compared
with the current results or with those of \cite{NBBMT13}. The results
of \cite{HRY16} are more in line
with ours, but the degree of temporal variations is indicated
only in passing\footnote{\cite{HRY16} state that during
  the last 200 days the mean magnetic field energy in the simulation
  `High' is almost 2.5 times higher than in the time average over the
  full duration (50~years) of the simulation.} and it is not possible
to assess the steepness of the increasing trend of the energy of the
mean field as a function of $\ReM$ accurately.

\section{Conclusions}

We find that, as the SGS Prandtl number (responsible for subgrid scale
turbulent transport) is increased, the rotation profiles realized in
the simulations develop a region of negative shear at mid-latitudes.
At moderate $\ReM$ this latitude coincides with a transition from poleward
migrating to equatorward migrating dynamo modes. This can be explained
by interpreting the solutions as dynamo waves propagating along the
isocontours of constant shear \citep{Pa55a,Yo75}. However, it appears
that, as $\PrM$ is sufficiently high (corresponding to high $\ReM$), the
regular cycles give way to
quasi-stationary or irregularly varying solutions.
However, due to computational constraints, the time series of these
runs are typically significantly shorter than of the lower--$\ReM$ runs, so
cyclic solutions with long cycle periods cannot be ruled out.

We also find that the cycle length of the dynamo solution undergoes a
systematic increase when the magnetic Prandtl number is increased: seemingly
independent of the other parameters, the decreasing magnetic diffusivity
leads to longer cycles. The highest Reynolds number runs are, however,
too short for our period analysis to work conclusively.

We find a strong dependence of the differential rotation on the
magnetic Reynolds number so that for the highest values of $\ReM$ both
radial and latitudinal shear are almost absent. The strongest
quenching tends to appear in cases where a small-scale dynamo is
present. However, there are exceptions.
The physical reason for the quenching is therefore unclear, but
several mechanisms appear plausible. First, the small-scale magnetic field
can enhance turbulent viscosity or quench the $\Lambda$-effect
responsible for maintaining differential rotation.
Suppression of {\em
  small-scale} flows due to the small-scale magnetic fields has been
suggested by \cite{HRY16}.
We find that at intermediate $\ReM$, the Maxwell stress is of
  the same order of magnitude as the Reynolds stress and becomes
  comparable at high $\ReM$. 
  Because of their opposite signs, the total stress is diminished.
  We expect that for intermediate and in particular for high $\ReM$,
  the Maxwell stress plays an important role in the angular momentum transport.
Second, the dependence of turbulent
transport on the large-scale magnetic field is $\ReM$-dependent and can
lead to enhanced quenching in the parameter regime studied here in
comparison to earlier more laminar simulations. However, for a limited
parameter range we find that the system vacillates between two states
where either the differential rotation is strong and mean magnetic field
relatively weak or vice versa. Furthermore, the temporal variations appear
to increase as the Rayleigh and Reynolds numbers are increased. Finally,
we find that in cases where the differential rotation is anti-solar,
the quenching is much less prominent, possibly due to significantly
weaker mean fields generated in those cases.

The total magnetic energy grows monotonically as a function of $\ReM$
in all of our runs. The energy of the azimuthally averaged
fields is in all cases consistent with {\em increasing or constant}
mean fields at high $\ReM$. This is apparently at odds with the
anelastic full-sphere simulations of \cite{NBBMT13} who found a
steeply declining trend with magnetic Reynolds number. However, their
relatively rapidly rotating ($\Omega=3\Omega_\odot$) simulations allow
also significant low-order non-axisymmetric modes to develop that will
not show up in azimuthal averaging. A fair comparison is thus not
possible without a proper assessment of the non-axisymmetric
contributions. Our results are more in line with those of \cite{HRY16}
who use a model rotating at the solar rate where the large-scale
fields are more clearly axisymmetric.

A possible source of discrepancies between the current and previous
studies is the use of wedge geometry. Rigorous comparisons between
wedges and fully spherical simulations have not been done so
far. However, runs with similar parameters (Rayleigh, Taylor, and
Prandtl numbers) produce results that are in qualitative agreement;
compare, e.g., the results of \cite{KKBMT10} regarding cyclic dynamos
with those of \cite{BMBBT11}. Similarly, the more recent simulations
showing equatorward migration \citep{KMB12,ABMT15} appear to support
the validity of the wedge approach. Furthermore, wedges with full
$2\pi$ extent in longitude produce dynamos dominated by non-axisymmetric
large-scale fields with azimuthal dynamo waves \citep{KMCWB13,CKMB14}
that are also routinely seen in anelastic full sphere simulations
\citep[e.g.][]{YGCR15}. Lastly, the transition from anti-solar to
solar-like differential rotation occurs at very similar Coriolis
numbers in a wide range of simulations, including wedges with
artificially increased luminosity and rotation rate \citep{GYMRW14}.
The various modeling approaches also use a range of different
subgrid-scale models and parameter values (see
Tables~\ref{tab:methods} and \ref{tab:values}
in Appendix~\ref{sec:comp}) and are still able to
reproduce similar large-scale phenomena.
These comparisons suggest that the current results obtained in wedges
are likely to be realized in fully spherical simulations in the same
parameter regime.

Our final conclusion is that the current simulations are not near an
asymptotic regime where the large-scale results would be independent
of the microphysical diffusion coefficients. This is most strikingly
demonstrated by the steep quenching of the differential rotation and
the disappearance of regularly oscillating large-scale magnetic fields
at high values of $\ReM$.

\begin{acknowledgements}
  The authors thank an anonymous referee for useful comments and
  Paul Charbonneau and Hideyuki Hotta for providing information
  regarding their models.
  The simulations were performed using the supercomputers hosted by
  CSC -- IT Center for Science Ltd.\ in Espoo, Finland, who are
  administered by the Finnish Ministry of Education and in the HLRS
  supercomputing center in Stuttgart, Germany through the PRACE
  allocation `SOLDYN.' Financial support from
  the Academy of Finland grants No.\ 136189, 140970, 272786 (PJK), and
  272157 to  the ReSoLVE Centre of Excellence (PJK, MJK), as well as the
  Swedish Research Council grants 621-2011-5076 and 2012-5797, and the
  Research Council of Norway under the FRINATEK grant 231444
  are acknowledged.
  JW acknowledges funding by the Max-Planck/Princeton Center for
  Plasma Physics and funding from the People Programme (Marie Curie
  Actions) of the European Union's Seventh Framework Programme
  (FP7/2007-2013) under REA grant agreement No.\ 623609.
\end{acknowledgements}

\bibliographystyle{aa}
\bibliography{../bibtex/bib}


\appendix

\section{Comparison to other simulation methods}

\label{sec:comp}

The purpose of this appendix is to
compare the diffusion schemes and estimates of the Prandtl number in
the present study with several
methods presented in the literature.
We consider the papers by \cite{NBBMT13} and \cite{HRY16} that are
discussed in the main text as well as the study of \cite{GCS10} that
represents another established simulation method; see
Tables~\ref{tab:methods} and \ref{tab:values}
for details.

Estimating the Prandtl number from simulations using the solar
luminosity requires that the value of the radiative diffusivity in the
solar convection zone is known. For the following we use an
estimate of this quantity at $r=0.85R_\odot$. Standard solar
models \citep[e.g.][]{Stix02} indicate that radiation carries roughly
ten per cent of the flux at this radius such that $L^{\rm
  rad}\approx0.1L_\odot$. Using \Eq{equ:Frad} we can write
\begin{equation}
L^{\rm rad} = 4\pi r^2 F^{\rm rad} = - 4\pi r^2 K \pd_r T. 
\end{equation}
We then use the definition of the heat conductivity $K=\cP \rho \chi$,
assume that the temperature gradient is close to 
the adiabatic one, $\pd_r T \approx -g/\cP$, and insert
$g=GM_\odot/r^2$ to obtain
\begin{equation}
\chi = \frac{L^{\rm rad}}{4\pi G M_\odot \rho}.
\label{equ:radchi}
\end{equation}
Using the values quoted above and $\rho\approx50$~kg~m$^{-3}$
\citep{Stix02} we find that $\chi\approx 5 \cdot
10^{2}$~m$^2$~s$^{-1}$ for the Sun at $r=0.85R_\odot$.

The solar luminosity is adopted in the study of \cite{NBBMT13},
yielding a radiative diffusion coefficient of the order of the
estimate \Eq{equ:radchi}. They, however, do not present a detailed
description of their model but refer to \cite{BBBMT10} where the
corresponding quantity is denoted as $\kappa_r$. Furthermore, the
values of $\nu$ at
mid-convection zone depth are in the range $6\ldots13 \cdot
10^7$~m$^2$~s$^{-1}$ for their Cases D3[a,b,pm1,pm2] and a value of $2
\cdot 10^6$~m$^2$~s$^{-1}$ is estimated for Case S3 with a Smagorinsky
SGS model. These yield Prandtl numbers of the order of $10^5$ and
$10^4$, respectively. For the model `Low' of \cite{HRY16}, only the
surface value $\nu=10^8$~m$^2$~s$^{-1}$ is given. As $\nu$ is
proportional to $\rho^{-1/2}$, the value at mid-convection zone depth is a
few times $10^7$~m$^2$~s$^{-1}$ and the Prandtl number is of the order of
$10^5$. For their cases `Medium', `High', and `High-S', a slope-limited
diffusion scheme is used, so estimating the diffusion
coefficients is non-trivial. 
However, they compute the Reynolds number
(hereafter ${\rm Re}_{\rm SL}$) based on the Taylor microscale that
was obtained from kinetic energy spectra. They find that the
Reynolds numbers for `High' and `High-S' are roughly an order of
magnitude larger than for `Low', whereas for `Low' and `Medium'
they are comparable. Assuming that the turbulent velocities in all
cases are of similar strengths, we can estimate `slope-limited diffusion
viscosities' from $\nu_{\rm SL} = \nu_{\rm Low}\Rey_{\rm Low}/\Rey_{\rm
  SL}$. We thus infer that the value of the Prandtl number in the model
`Medium' is $10^5$ and an order of magnitude smaller in the models `High' and
`High-S'. In these runs, the surface value of the turbulent heat conductivity
is $\kappa=2\cdot10^9$~m$^2$~s$^{-1}$ (H.\ Hotta, private communication),
yielding $\PraSGS$ in the range $0.05\ldots2\cdot10^{-3}$ for the runs
of \cite{HRY16}.

In principle even a third Prandtl number can be defined
based on the diffusion coefficient applied only to the mean
(spherically symmetric) entropy profile. In the present study and also
in that of \cite{HRY16}, this coincides with the coefficient relevant
for diffusing entropy fluctuations. However, the coefficients can also
be different which is the formulation often used in simulations
performed with the Anelastic Spherical Harmonic (ASH) code including
those presented in
\cite{NBBMT13}. However, the value of the coefficient for mean entropy
diffusion ($\kappa_0$) is not provided in the reference with a
more detailed model description \citep{BBBMT10}.

The Eulerian-Lagrangian (EULAG) code of \cite{SC13}
employed by \cite{GCS10} uses quite a different approach
and replaces radiative conduction by Newtonian cooling toward a
prescribed thermodynamic state. The timescale of the cooling
$\tau_{\rm c}$ is typically of the order of 20 years
\cite[P.\ Charbonneau, private communication; see
  also][]{SBBCMS16}. This timescale would correspond to a
radiative/SGS diffusion coefficient of $\chi=\Delta r^2/\tau_{\rm c}
\approx 5\cdot 10^7$~m$^2$~s$^{-1}$, where $\Delta r = 0.25R_\odot$.
Although this comparison yields some insight about the entropy
evolution, the cooling and diffusion processes cannot be directly
equated. Therefore the concept of
a Prandtl number does not appear suitable in that case. The diffusion of
velocity, magnetic fields, and entropy fluctuations is due to the
numerical scheme making the estimates of the other Prandtl numbers
also problematic. Presumably the diffusivities of all variables at a given
resolution are roughly similar such that $\PraSGS$ and $\PrM$ are
of the order of unity. Hydrodynamic EULAG models with an
otherwise similar setup to that of \cite{GCS10} yield estimates of
viscosity and entropy diffusion in ranges $\nu_{\rm
  eff}=0.6\ldots1.2\cdot 10^8$~m$^2$~s$^{-1}$ and $\kappa_{\rm
  eff}=1\ldots 8\cdot 10^7$~m$^2$~s$^{-1}$, respectively, with
$\PraSGS=\nu_{\rm eff}/\kappa_{\rm eff}\approx1\ldots8$
\citep{SBBCMS16}.

We show the conversion factors for the definitions used in the current study
for P\'eclet, fluid and magnetic Reynolds, and Coriolis numbers in
\Table{tab:values}. The conversion for the results of \cite{NBBMT13}
is straightforward, except that for the Coriolis number we have used
the relation $\Ta=\Co'^2\Rey'^2$ to compute $\Co'=2\pi\Co$. The
definition of the Rossby number in \cite{NBBMT13},
$\Ro=\omega/2\Omega$, is based on the vorticity
$\bm\omega=\bm\nabla\times\bm{u}$ and corresponds to
$\Co=\Ro^{-1}k_\omega/\kf$, where $k_\omega=\omega_{\rm
  rms}/\urms$. The values of $\Co$ for the models of
\cite{NBBMT13} correspond well to our simulations with
$\Omega=3\Omega_\odot$, see \Table{tab:runs}. In the case of the EULAG
simulations we use the values quoted by \cite{PC14} for the Reynolds
numbers ($30\ldots60$ in their notation) divided by $2\pi$ and the
estimate of $\PraSGS$ quoted above to compute $\Pe$. The estimate of
the Coriolis number is based on simulation data provided by
P.\ Charbonneau. Finally, for the study of \cite{HRY16}, the conversion
of the fluid and magnetic Reynolds numbers is based on values given in
the paper whereas the estimates of $\Pe$ and $\Co$ are based on
simulation data provided by H.\ Hotta. The values of $\Co$ are again
in excellent agreement with our corresponding simulations with
$\Omega=\Omega_\odot$; see \Table{tab:runs}.

We note that the simulations of \cite{NBBMT13} and \cite{HRY16}
operate in a low-$\PraSGS$ regime which is also realized in the Sun,
although in a much more extreme fashion. Such a regime is required
especially in simulations with solar luminosity and rotation rate to
lower the convective velocities and to achieve solar-like differential
rotation \citep[][]{KKB14,HRY16}. However, the tradeoff is that the
P\'eclet numbers are low and the evolution of entropy is significantly
influenced by the SGS diffusion. It is not clear whether such an
approach is more realistic in comparison to having $\PraSGS$ and
$\PrM$ of the order of unity and $\Rey$, $\ReM$, and $\Pe$ $\gg1$. We
also note that the run times of the highest resolution runs are short:
4 years for `S3' in \cite{NBBMT13} and 500 days for `High-S' in
\cite{HRY16}, whereas transients and/or secular evolution in the
high-$\ReM$ regime can have a significantly longer timescale; see
\Fig{fig:penet}(b)-(d).

\begin{table*}[t!]
\centering
\caption[]{Diffusion schemes applied in a few comparable studies.}
  \label{tab:methods}
      $$
          \begin{array}{ccccccccccc}
          \hline
          \hline
          \noalign{\smallskip}
          \mbox{Study} & \mbox{Viscosity} & \mbox{Magnetic} & \mbox{Radiative} & \mbox{Mean entropy} & \mbox{Fluctuating entropy} & \\
           &  & \mbox{diffusivity} & \mbox{diffusion} & \mbox{diffusion} & \mbox{diffusion} \\
          \hline
          \hline
          \mbox{Present work} & \nu & \eta & \chi & \chiSGS & \chiSGS & \\ 
          \mbox{({\sc Pencil Code})} & (\mbox{constant}) & (\mbox{constant}) & (6.4\cdot10^5 \times \mbox{solar})  & (\mbox{piecewise constant}) & (\mbox{piecewise constant}) & \\
          \hline
          \hline
          \mbox{\cite{NBBMT13}, (ASH)} & \nu & \eta & \kappa_r & \kappa_0 & \kappa & \\ 
          \mbox{Cases D3[a,b,pm1,pm2]} & (\propto \rho^{-1/2}) & (\propto \rho^{-1/2}) & (\mbox{solar}) & (\mbox{near surface}) & (\propto \rho^{-1/2}) & \\
\hline
          \mbox{\cite{NBBMT13},} & \nu_S & \eta_S\equiv\nu_S/{\rm Pm} & \kappa_r & \kappa_0 & \kappa_S\equiv\nu_S/\Pra & \\
          \mbox{Case~S3} & \mbox{Smagorinsky} & \mbox{Smagorinsky}  & (\mbox{solar}) & (\mbox{near surface}) & \mbox{Smagorinsky} & \\
          \hline
          \hline
          \mbox{\cite{GCS10}} & \mbox{iLES} & \mbox{iLES} & \mbox{Newtonian} & \mbox{Newtonian} & \mbox{iLES+Newtonian} & \\ 
          \mbox{(EULAG)} &  &  & \mbox{cooling} & \mbox{cooling} & \mbox{cooling} & \\ 
          \hline
          \hline
          \mbox{\cite{HRY16}} & \nu & \eta & \kappa_r & \kappa & \kappa & \\
          \mbox{Low} & (\propto \rho^{-1/2}) & (\propto \rho^{-1/2}) & (\mbox{solar}) & (\propto \rho^{-1/2}) & (\propto \rho^{-1/2}) & \\ 
          \hline
          \mbox{\cite{HRY16}} & \mbox{Slope-limited} & \mbox{Slope-limited} & \kappa_r & \kappa & \kappa & \\ 
          \mbox{Medium, High, High-S} & \mbox{diffusion} & \mbox{diffusivity} & (\mbox{solar}) & (\propto \rho^{-1/2}) & (\propto \rho^{-1/2}) & \\ 
          \hline
          \hline
          \end{array}
          $$ \tablefoot{The entries for each study correspond to the
              symbols used for the various diffusion coefficients and
              their spatial profiles or functional dependences, where
            applicable. The coefficients $\kappa_r$ and $\kappa_0$ are
            not mentioned in the paper by \cite{NBBMT13} who refer to
            an earlier study by \cite{BBBMT10} for a detailed
            description of their model. iLES stands for implicit
            Large-Eddy
            Simulation where the truncation errors of the numerical
            scheme provide the diffusion.}
\end{table*}

\begin{table*}[t!]
\centering
\caption[]{Parameter values with conversion factors to the
    definition in the present study from the same studies as in
    \Table{tab:methods}.}
  \label{tab:values}
      $$
          \begin{array}{cccccccccccccc}
          \hline
          \hline
          \noalign{\smallskip}
          \mbox{Study} & \Pra & \PraSGS & \PrM & {\rm Pe} & \Rey & \ReM & \Co \\
          \hline
          \hline
          \mbox{Present work} & 18\ldots72 & 0.25\ldots5 & 0.5\ldots5 & 4 \ldots 134 & 17\ldots134 & 5 \ldots 151 & 1.4 \ldots 15.9 \\
          \hline
          \hline
          \mbox{\cite{NBBMT13}, (ASH)} & \nu/\kappa_r & \nu/\kappa & \nu/\eta & \Pra\Rey'/(2\pi) & \Rey'/(2\pi) & \Rm'/(2\pi) & \sqrt{\Ta}/(2\pi\Rey') \\
          \mbox{Cases D3[a,b,pm1,pm2]} & \mathcal{O}(10^5) & 0.25  & 0.5\ldots2 & 4 \ldots 11 & 16 \ldots 43 & 8 \ldots 22 & 4.5\ldots5.5 \\
\hline
          \mbox{\cite{NBBMT13},} & \nu_S/\kappa_r & \nu_S/\kappa_S & \nu_S/\eta_S & \Pra\Rey'/(2\pi) & \Rey'/(2\pi) & \Rm'/(2\pi) & \sqrt{\Ta}/(2\pi\Rey') \\
          \mbox{Case~S3} & \mathcal{O}(10^4) & 0.25  & 0.5 & 229 & 915 & 458 & 5.8 \\
          \hline
          \hline
          \mbox{\cite{GCS10}} & - &  - & - & - & - & - & - \\
          \mbox{(EULAG)} & ? & \approx 1 & \approx 1 & 5\ldots10 & 5\ldots10 & 5\ldots10 & 30 \\ 
          \hline
          \hline
          \mbox{\cite{HRY16}} & \nu/\kappa_r & \nu/\kappa & \nu/\eta & \Pr\Rey/(2\pi) & \Rey/(2\pi) & \Pm\Rey/(2\pi) & - \\
          \mbox{Low} & \mathcal{O}(10^5) & 0.05 & 1 & 2.5 & 51 & 51 & 1.4 \\
          \hline
          \mbox{\cite{HRY16}} & \nu_{\rm SL}/\kappa_r & \nu_{\rm SL}/\kappa & \nu_{\rm SL}/\eta_{\rm SL} & \Pr\Rey/(2\pi) & \Rey/(2\pi) & \Pm\Rey/(2\pi) & - \\
          \mbox{Medium, High, High-S} & \mathcal{O}(10^4\ldots10^5) & 0.04\ldots2\cdot10^{-3} & 1.03\ldots1.71 & 2.5 & 61 \ldots 1129 & 63\ldots1930 & 1.4 \\
          \hline
          \hline
          \end{array}
          $$ \tablefoot{Question marks indicate
            quantities not described in the respective papers.
            In the last row, $\nu_{\rm SL}$ and $\eta_{\rm SL}$ are
            the viscosity and magnetic diffusivity estimated from their
            slope-limited diffusion scheme (see text).
            The values for $\Pe$, $\Rey$, and
            $\ReM$ for the study of \cite{GCS10} are based on values
            stated in \cite{PC14} divided by $2\pi$. The Coriolis
            numbers for the simulations of \cite{GCS10} and
            \cite{HRY16} were calculated using our definition,
            \Eq{eq:Coriolis}, and simulation data for the volume
            averaged rms-value of velocity fluctuations provided by
            P.\ Charbonneau and H.\ Hotta, respectively.}
\end{table*}

\end{document}